\documentclass[12pt,pdflatex,sn-nature]{sn-jnl}

\usepackage{ragged2e}
\usepackage{graphicx}
\usepackage{multirow}
\usepackage{amsmath,amssymb,amsfonts}
\usepackage{amsthm}
\usepackage{mathrsfs}
\usepackage[title]{appendix}
\usepackage{xcolor}
\usepackage{textcomp}
\usepackage{manyfoot}
\usepackage{booktabs}
\usepackage{algorithm}
\usepackage{algorithmicx}
\usepackage{algpseudocode}
\usepackage{listings}

\hypersetup{hidelinks,pdftitle={Scalable quantum error mitigation for dynamical decoupling via Hadamard phase cycling},pdfauthor={Weibin Ni, Zhijie Li, Guanyu Qu, Asif Equbal, Zhecheng Sun, Jiale Dai, Fazhan Shi, Lei Sun}}

\begin{document}
\begingroup

\title[Scalable quantum error mitigation for dynamical decoupling via Hadamard phase cycling]{\centering Scalable quantum error mitigation for dynamical decoupling via Hadamard phase cycling }

\author[1,2,3]{\fnm{Weibin} \sur{Ni}}
\author[4]{\fnm{Zhijie} \sur{Li}}
\author[4]{\fnm{Guanyu} \sur{Qu}}
\author[5,6]{\fnm{Asif} \sur{Equbal}}
\author[3,7]{\fnm{Zhecheng} \sur{Sun}}
\author[2,3]{\fnm{Jiale} \sur{Dai}}
\author[4,8]{\fnm{Fazhan} \sur{Shi}}
\author[2,3,7,*]{\fnm{Lei}\sur{Sun}}\email{sunlei@westlake.edu.cn}

\makeatletter
\gdef\corrauthemail{sunlei@westlake.edu.cn}
\makeatother

\affil[1]{\orgdiv{Department of Physics}, \orgname{Fudan University}, \orgaddress{\city{Shanghai} \postcode{200433}, \country{China}}}

\affil[2]{\orgdiv{Department of Physics, School of Science and Research Center for Industries of the Future}, \orgname{Westlake University}, \orgaddress{\city{Hangzhou} \postcode{310030}, \state{Zhejiang}, \country{China}}}

\affil[3]{\orgdiv{Institute of Natural Sciences}, \orgname{Westlake Institute for Advanced Study}, \orgaddress{\city{Hangzhou} \postcode{310024}, \state{Zhejiang}, \country{China}}}

\affil[4]{\orgdiv{Laboratory of Spin Magnetic Resonance, School of Physical Sciences, Anhui Province Key Laboratory of Scientific Instrument Development and Application}, \orgname{ University of Science and Technology of China}, \orgaddress{\city{Hefei}, \postcode{230026}, \state{Anhui}, \country{China}}}

\affil[5]{\orgdiv{Center for Quantum and Topological Systems},
	\orgname{New York University Abu Dhabi},
	\orgaddress{\city{Abu Dhabi}, \postcode{129188}, \country{United Arab Emirates}}}

\affil[6]{\orgdiv{Department of Chemistry},
	\orgname{New York University Abu Dhabi},
	\orgaddress{\city{Abu Dhabi}, \postcode{129188}, \country{United Arab Emirates}}}

\affil[7]{\orgdiv{Department of Chemistry, School of Science and Research Center for Industries of the Future},
	\orgname{Westlake University},
	\orgaddress{\city{Hangzhou}, \postcode{310030}, \state{Zhejiang}, \country{China}}}

\affil[8]{\orgdiv{School of Biomedical Engineering and Suzhou Institute for Advanced Research},
	\orgname{University of Science and Technology of China},
	\orgaddress{\city{Suzhou}, \postcode{215123}, \country{China}}}

\maketitle

\section*{Abstract}
Quantum coherence remains a fundamental challenge for advancing quantum technologies. Although dynamical decoupling can suppress decoherence noise,    control errors can cause the overestimation of decoherence times. Here, we present Hadamard phase cycling, a scalable non-Markovian quantum error mitigation method using group-structured phase configurations to    select target coherence pathways. Validated across molecular electron spins, nitrogen-vacancy centers in diamond, nuclear spins, trapped ions, and superconducting qubits, this technique enables accurate decoherence time characterization and enhanced state fidelity with linear complexity. Our results    show that control-induced  coherence-population mixing    can substantially prolong measured echo decays in the systems studied. This mechanism provides grounds to question decoherence times obtained under comparable conditions when population-mediated contributions have not been excluded. By selecting target coherence pathways, Hadamard phase cycling establishes a robust framework for reliable quantum control, paving the way for reassessment and advancement of coherence benchmarks in the NISQ era.

\section*{Introduction}

Quantum information science holds great promise for delivering transformative computational power, metrological sensitivity, and communication security beyond classical limits\cite{ref1}. In the current Noisy Intermediate-Scale Quantum (NISQ) era, the realization of quantum advantages is hampered by susceptibility of qubits to decoherence and control errors\cite{ref2}. In principle, quantum error correction can eliminate these errors and enable fault-tolerant quantum technologies, but it imposes stringent demands on error rates and requires a substantial overhead of physical qubits, conditions that are presently beyond technological reach\cite{ref3,ref4}.

Dynamical decoupling, which applies a pulse sequence to average out unwanted environment-system coupling, is a well established strategy to suppress the decoherence error\cite{ref5,ref6,ref7}. It has been widely applied to extend decoherence times ($T_2$) for versatile qubit platforms, e.g. nuclear spins\cite{ref8}, solid-state spin defects\cite{ref9,ref10,ref11,ref12}, neutral atoms\cite{ref13}, superconducting circuits\cite{ref14,ref15}, trapped ions\cite{ref16}, semiconductor quantum dots\cite{ref17,ref18}, paramagnetic molecules\cite{ref19,ref20,ref21,molecule_addition}, etc. The capability of coherence preservation enables applications of dynamical decoupling across a wide range of quantum technologies, including Hamiltonian engineering in quantum simulation\cite{ref22}, noise spectrum reconstruction\cite{ref23,ref24}, sensitive and selective quantum metrology\cite{ref25}, and state protection in quantum computing\cite{ref26}. Meanwhile, it is also widely applied in nuclear magnetic resonance (NMR), electron paramagnetic resonance (EPR), and magnetic resonance imaging (MRI), empowering precise measurements of spin-spin interaction as well as the fast spin echo method for efficient MRI\cite{ref27}.

However, dynamical decoupling itself is prone to control errors, causing accumulation of erroneous dynamic evolution of quantum states that leads to low state fidelity. Although some dynamical decoupling sequences, exemplified by CPMG and XY-8, are designed with inherent robustness to protect the fidelity of final state\cite{ref28}, control errors may still reduce the fidelity of intermediate states and compromise characterization of dynamic properties\cite{ref29}. Therefore, albeit widely used for acquiring $T_2$, these robust sequences tend to misestimate $T_2$ under control errors.    Such biases can affect qubit  design and optimization  when control-error-induced contributions are not distinguished from the target coherence signal \cite{ref30}. Moreover, control errors may introduce spurious signals in quantum sensing\cite{spurious} and distort echo shapes in fast spin-echo MRI,    complicating signal interpretation. Addressing these challenges demands accurate and efficient quantum error mitigation (QEM) tailored for dynamical decoupling.

 Here, we show that for systems whose relaxation time ($T_1$) well exceeds $T_2$ and where single gate fidelity is relatively low, dynamical decoupling without proper mitigation of control errors tends to severely overestimate $T_2$ or become impractical to implement. Complete error mitigation demands exponential scaling with the circuit depth according to the selection rule of coherence-only dynamic processes, yet it can be well approximated with linear scaling by quantum circuits whose phase configurations form an abelian group. Based on this property, we design Hadamard phase cycling, the first practical QEM method for inversion-pulse-based dynamical decoupling (IDD), e.g. CP, CPMG, XY-8, XY-16, and UDD, which are widely used in quantum information science and magnetic resonance spectroscopy. Applying this protocol to IDD enables accurate measurement and significant enhancement of $T_2$ for ensemble molecular electron spin qubits, nitrogen-vacancy (NV) centers in diamond, and nuclear spin qubits. Benchmarking it on single trapped $^{40}$Ca$^+$ ion and superconducting transmon qubits yields near-quantitative effective state fidelity.

\section*{Results}
\subsection*{Practical phase cycling for IDD}
QEM is a highly feasible strategy to handle control errors of pulses (i.e. single quantum gates)\cite{ref31,ref32}. By post-processing outputs from an ensemble of noisy quantum circuits, QEM reduces the impact of errors and more accurately estimates expectation values of observables. Phase cycling, which was developed decades ago in the context of NMR and then extended to EPR\cite{ref33}, MRI\cite{ref34}, and optical spectroscopy\cite{ref35}, is likely the first QEM method. Typically, phase cycling employs a set of functionally equivalent quantum circuits (i.e. pulse sequences) leveraging their phase degree of freedom. The phase configurations are systematically designed by classifying qubit dynamics based on their evolution pathways\cite{ref36}. This approach selectively extracts the desired dynamics by averaging observables obtained from all circuits. Hence, phase cycling is a non-Markovian QEM yet both its potential in quantum information science and its application for dynamical decoupling have remained largely unexplored.

We consider a qubit in a dephasing environment under IDD with systematic control errors (Fig. 1a). Rapid environmental fluctuations beyond the refocusing capability of the control pulses are captured as an exponential decay. To describe the resulting qubit dynamics, the quantum state is decomposed into coherence orders $p \in \{0, \pm1\}$, where $p = \pm1$ represents transverse coherence subject to decoherence and $p = 0$ denotes longitudinal population subject to relaxation. Each control pulse acts as a branch point (Fig. 1e), redistributing these orders with weights determined by control imperfections. Consequently, a multi-pulse sequence generates an exponentially vast manifold of coherence transfer pathways, each encoding distinct non-Markovian dynamics.

Reliable implementation of phase cycling could construct an effective quantum channel that selectively isolates the target dynamics from undesired pathways. However, most previous studies implemented IDD without phase cycling or with two-step phase cycling (TPC). TPC suffices to select desired dynamic processes only when all pulses are error-free, whereas noisy inversion pulses induce unintended coherence-order changes ($\Delta p=\pm 1$) and cause coherence-population mixing (Fig. 1b,e). The pathways consisting of only $p=\pm 1$ states generate desired signals reflecting pure decoherence; those involving $p = 0$ states generate undesired signals that undergo intermittent relaxation processes\mbox{\cite{stimulatedEcho}}. To illustrate this point, we applied a modified CPMG sequence on an ensemble of Cu$^{2+}$-based molecular qubits (Fig. 2b inset) to separate desired and undesired spin echoes (Fig. 1c)\cite{ref37}. The former decays more than 600 times faster than the latter (Fig. 1d). In standard CPMG experiments, these two types of echoes overlap, causing TPC to overestimate $T_2$ that leads to positively biased thereby misleading assessment of qubit performance.

Rigorous QEM for IDD requires precise isolation of coherence-only dynamic processes. In an IDD-$m$ sequence, imperfections in any combination of the $m$ pulses can trigger transitions between coherence and population. Consequently, the signal becomes a superposition of desired decoherence pathways and an exponentially large manifold of undesired pathways involving relaxation ($2^m$ possibilities). We extract the former by exploiting the phase degree of freedom: applying a $\pi$ phase shift to a pulse inverts the sign of any signal whose corresponding pathway involves an error at that pulse (Fig. 1c). Accordingly, we represent a pulse phase by a binary basis $\left \{\pm1 \right \}$, where $+1$ denotes the original phase and $-1$ denotes a $\pi$ phase shift. In theory, eliminating undesired signals requires summing over a complete basis of phase configurations for all pulses, a method termed as complete phase cycling (CPC)\cite{ref19,ref30,ref38}.  Although the CPC and its exponential cost are well known, to the best of our knowledge the previous studies do not provide an explicit lower-bound proof for the complete binary pathway-selection problem considered here. We prove that complete selection requires at least $2^m$ phase cycling steps (Supplementary Information).

We surmount this exponential barrier by approximating the filtering power of CPC while implementing linear scaling. Mathematically, phase configurations forming an abelian group ensure destructive interference and vanishment of most erroneous pathways upon averaging.  Accordingly, we design phase configurations using Sylvester-constructed Hadamard matrices\cite{ref39}, i.e. square matrix whose columns form an abelian group under Hadamard product, and integrate them with a parity-based selection rule, yielding the Hadamard phase cycling (HPC) protocol (Fig. S1; see details in the supplementary information):

\begin{equation}
	HPC=
	\begin{pmatrix}
		+ & +\vec{1}_m & +\vec{1}_m & +H'\\
		+ & +\vec{1}_m & +\vec{1}_m & -H' \\
		- & -\vec{1}_m & +\vec{1}_m & +H' \\
		- & -\vec{1}_m & +\vec{1}_m & -H'
	\end{pmatrix}.
\end{equation}
Here, the columns represent the phase settings for all pulses across different quantum circuits, $\vec{1}_m$ is an all-ones vector of size $m$, $H'$ is a Sylvester-constructed Hadamard matrix excluding the first column, and the sign column dictates the data processing (addition or subtraction). For a sequence of length $m$ (where $m$ is a power of 2 for simplicity), HPC reduces the     number of phase configurations (e.g., complexity)  from $2^m$ to $4m$. In the combinatorial counting of erroneous pathway classes, it  filters more than 98$\%$   for $m>16$ (Fig. S2).  This fraction counts pathway classes; their contributions to the measured signal also depend on their amplitudes and subsequent dynamics.     Nevertheless, HPC suppresses a large fraction of undesired pathway classes with only a linear number of phase cycling steps, providing a scalable approximation to CPC (Fig. 2a). HPC also provides a practical reference for testing experimentally whether TPC is sufficient for a given measurement.

\begin{figure}[!htbp]
		\centering
		\includegraphics[width=\textwidth]{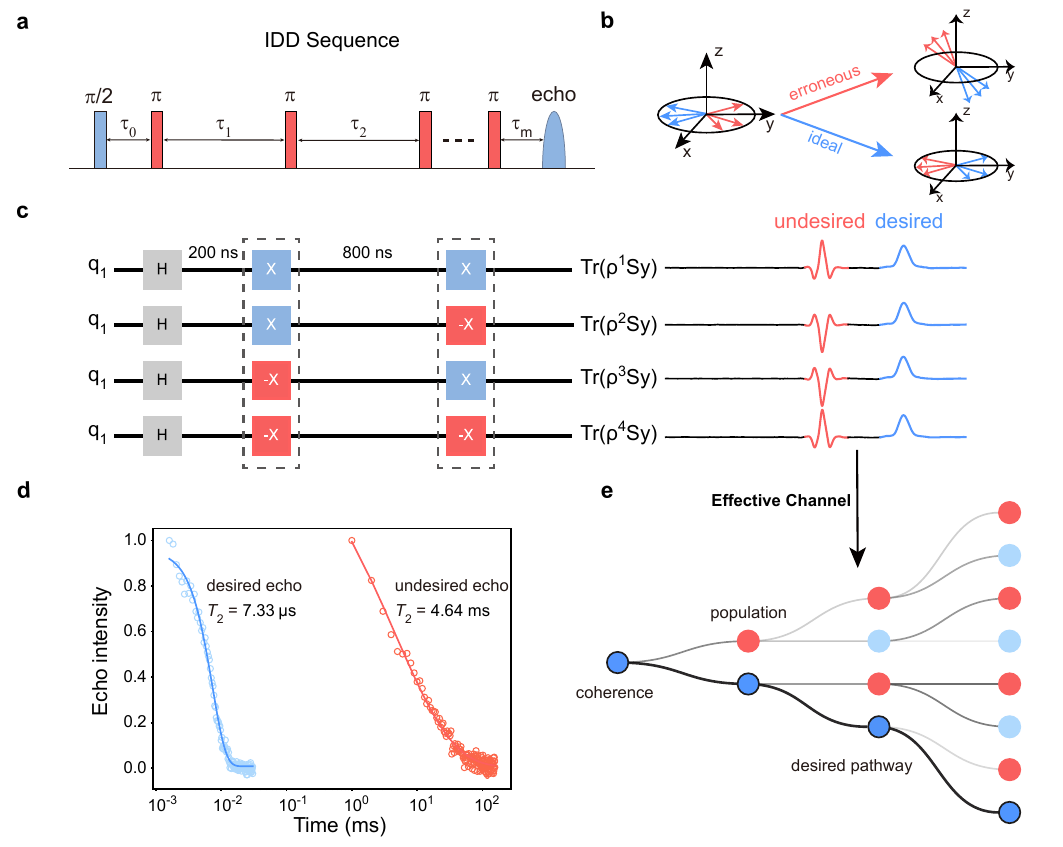}

		\caption{\textbf{Introduction to phase cycling.}
			\textbf{a}, Composition of a general IDD sequence. For the CPMG sequence, the $\pi$ pulses are applied at constant time intervals. Each $\pi$ pulse has a phase that differs by $\pi/2$ from the phase of the initial $\pi/2$ pulse. \textbf{b}, Illustrative figure of desired and undesired dynamical processes of qubits under noisy IDD sequence. \textbf{c}, CPC for modified CPMG-2. q$_1$ and H represent qubit and Hadamard gate, respectively. $\pm X$ represent inversion pulses with opposite phases. $Tr(\rho^i S_y)$ represents the expectation value of $S_y$ in the final state $\rho^i$ of the $i^{th}$ quantum circuit. Schemes on right are experimental outputs of quantum circuits with desired and undesired echoes highlighted. \textbf{d}, Echo decay curves of desired and undesired echoes. Data were collected for ensemble Cu$^{2+}$-based molecular qubits at 8 K by varying the interval between two inversion pulses. The desired and undesired echoes decay with apparent $T_2$ of 7.33 $\mu s$ and 4.64 ms, respectively. \textbf{e}, This branching tree illustrates the exponential proliferation of coherence transfer pathways, where each imperfect control pulse redistributes the quantum state between coherence (blue) and population (red) manifolds. Reliable phase cycling selects desired coherence-only pathways.}
		\label{fig:1}
	\end{figure}

\subsection*{Quantum error mitigation for ensemble qubits}
The non-Markovian nature of HPC enables its application for accurate $T_2$ acquisition. As proof of concept, we implemented Hadamard, complete, and two-step phase cycling for CPMG-$m$ to acquire $T_2$ of the ensemble Cu$^{2+}$-based molecular qubit (Fig. 2b inset) with pulse EPR spectroscopy. We first fixed the delay time between adjacent pulses while increasing the number of pulses. Hadamard and complete phase cycling yielded nearly identical echo intensities for each sequence with $ m$ = 2 -- 16 (Fig. 2b). This phenomenon reinforces the nearly quantitative error mitigation of HPC and shows the advantage of its linear scalability: for $m=16$, HPC exerts 64 quantum circuits while CPC demands 65536 circuits. In sharp contrast, TPC leads to a much slower echo decay with increasing $m$, giving rise to significantly stronger echoes (Fig. 2b). This indicates substantial coherence-population mixing: TPC on noisy quantum circuits does not probe pure decoherence and more reliable phase cycling is indispensable to mitigate control errors.

The distinction between two-step and Hadamard phase cycling becomes critical when measuring $T_2$ with fixed number of pulses and variable delay times. Both methods exhibit an increase of the apparent $T_2$ with increasing sequence length, yet they display drastically different scaling factors: $T_2\propto m^{0.42}$ for TPC and $T_2\propto m^{0.17}$ for HPC (Fig. 2c). CPMG with TPC requires 256 pulses to improve $T_2$ by more than 10 times, whereas HPC would enforce nearly 0.9 million pulses should its scaling law be applicable to longer sequences. From another perspective, TPC would overestimate $T_2$ by an order of magnitude for CPMG-8192. Under proper error mitigation provided by HPC, CPMG-128 improves $T_2$ to $24~\mathrm{\mu s}$, approximately four times the value obtained from Hahn echo decay (Fig. S3), demonstrating utility of dynamical decoupling for coherence protection. Similar behavior were observed for ensemble radical-based molecular qubits embedded in a covalent organic framework\cite{ref40} and nuclear spins in $^{13}$C-labelled glycine (Fig. S4 and S5).

We observed similar phenomena in another representative ensemble electron spin qubit system, i.e. NV centers in diamond (Fig. 2d inset). For CPMG-$m$ experiments with fixed delay time, the echo intensity decays much faster under HPC than TPC as $m$ increases (Fig. 2d). For those with fixed sequence length and variable delay time, TPC gives rise to longer apparent $T_2$ than HPC for CPMG-32 and CPMG-64 (Fig. S6). Remarkably, employing CPMG-1536 with TPC yields an apparent $T_2$ approaching 0.25$T_1$, indicating that such a long sequence mixes decoherence with relaxation (Fig. 2e). Overall, these experiments confirm that although TPC generates seemingly more promising coherence enhancement, it inevitably involves coherence-population mixing and causes significant overestimation of $T_2$. HPC, which balances effectiveness and scalability for error mitigation, provides a practical solution for $T_2$ acquisition of ensemble qubit systems.

\begin{figure}[!htbp]
		\centering
		\includegraphics[width=\textwidth]{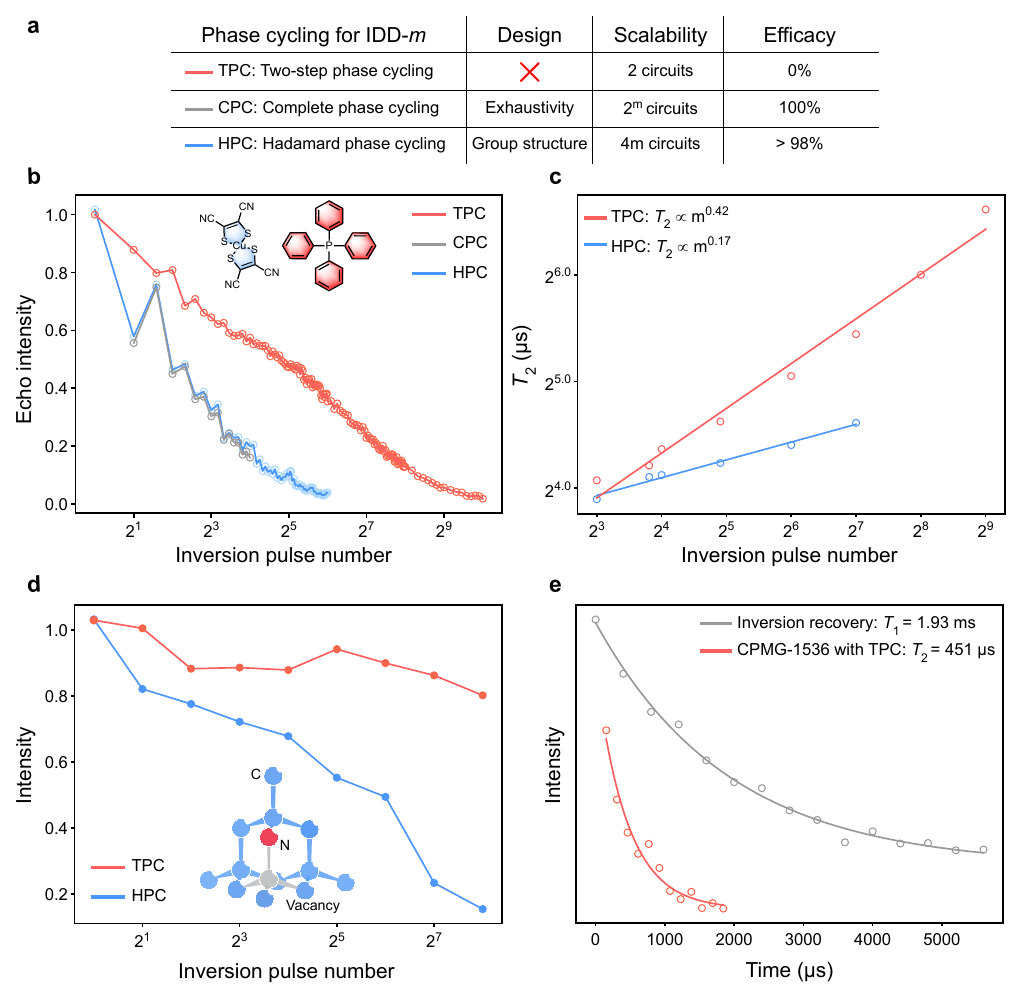}

		\caption{\textbf{$\textbf{T}_\textbf{2}$ measurements for ensemble electron spin qubits under CPMG.}
			\textbf{a}, Introduction of different phase cycling methods. Assume that $m$ is a power of 2 and $m > 16$ for which CPC becomes impractical. \textbf{b}, Comparison of echo intensities collected with HPC, CPC, and TPC. Inset: Schematic illustrations of the Cu$^{2+}$-based molecular qubit. \textbf{c}, Dependencies of $T_2$ on the number of inversion pulses for HPC and TPC. Data in \textbf{b} and \textbf{c} were collected on ensemble Cu$^{2+}$-based molecular qubits at 8 K. \textbf{d}, Comparison of echo intensities collected with HPC and TPC. Inset: Schematic illustrations of NV center in diamond. \textbf{e}, Inversion recovery curve and echo decay curve under CPMG-1536 with TPC. Data in \textbf{d} and \textbf{e} were collected on ensemble NV centers in diamond at room temperature. The x-axes of \textbf{b}-\textbf{d} and y-axis of \textbf{c} are in log scale.}
		\label{fig:2}
	\end{figure}

HPC is also applicable to UDD, which    has optimal error-suppression properties under the assumptions of its  pure-dephasing model\cite{ref41,ref42} (Fig. 3a). For the Cu$^{2+}$-based molecular qubits, complete and Hadamard phase cycling again yielded well-defined echoes with identical intensities for UDD-$m$ with $m$ = 2 -- 10 (Fig. 3b). In contrast, TPC generated complex echo signals with wrong phases likely due to overlapping of desired echoes with undesired ones (Fig. 3b inset). UDD-$m$ with HPC shows $T_2\propto m^{0.81}$ (Fig. 3c).    Under the same HPC protocol, the fitted $T_2$ increases more rapidly with pulse number for UDD than for CPMG, which has an exponent of  0.17, in this Cu$^{2+}$ sample at 8 K. Specifically, UDD-8 slightly outperforms CPMG-64 (Fig. 3d), and UDD-14 improves $T_2$ to $35~\mathrm{\mu s}$ that well exceeds the value observed for CPMG-128 ($T_2$ = $24~\mathrm{\mu s}$).    These scaling exponents characterize the sequence performance under the noise environment, control conditions, pulse-number ranges, and fitting procedure used here; they are not universal sequence constants or transferable predictions for other samples.

The TPC--HPC comparisons for a fixed sequence assess the effect of phase cycling, whereas the UDD--CPMG comparison under common HPC assesses the relative sequence performance in this sample. The distorted UDD echoes under TPC and the agreement between HPC and CPC for the tested short sequences demonstrate the utility of HPC for evaluating the target coherence response. The results for  periodic CPMG and    aperiodic UDD support the applicability of HPC to different IDD designs, with its practical benefit depending on the sequence, system, and measurement task.

For more robust DD sequences, e.g., XY-8 and XY-16 sequences, HPC provides a complementary means of mitigating control errors. Robust DD compensates pulse errors within the sequence\mbox{\cite{ref28}}, whereas HPC selects signal contributions through phase cycling. These different mechanisms may improve control robustness when combined (Supplementary Fig.~S9) and help suppress spurious responses in quantum sensing\mbox{\cite{spurious,HadamardSensingChannel}}. The broad applicability of HPC to IDD and its complementarity with robust sequences motivate its use in evaluating DD sequences and further applications in quantum sensing\mbox{\cite{ref25}} and fast spin-echo MRI\mbox{\cite{ref27}}.

\begin{figure}[!htbp]
		\centering
		\includegraphics[width=\textwidth]{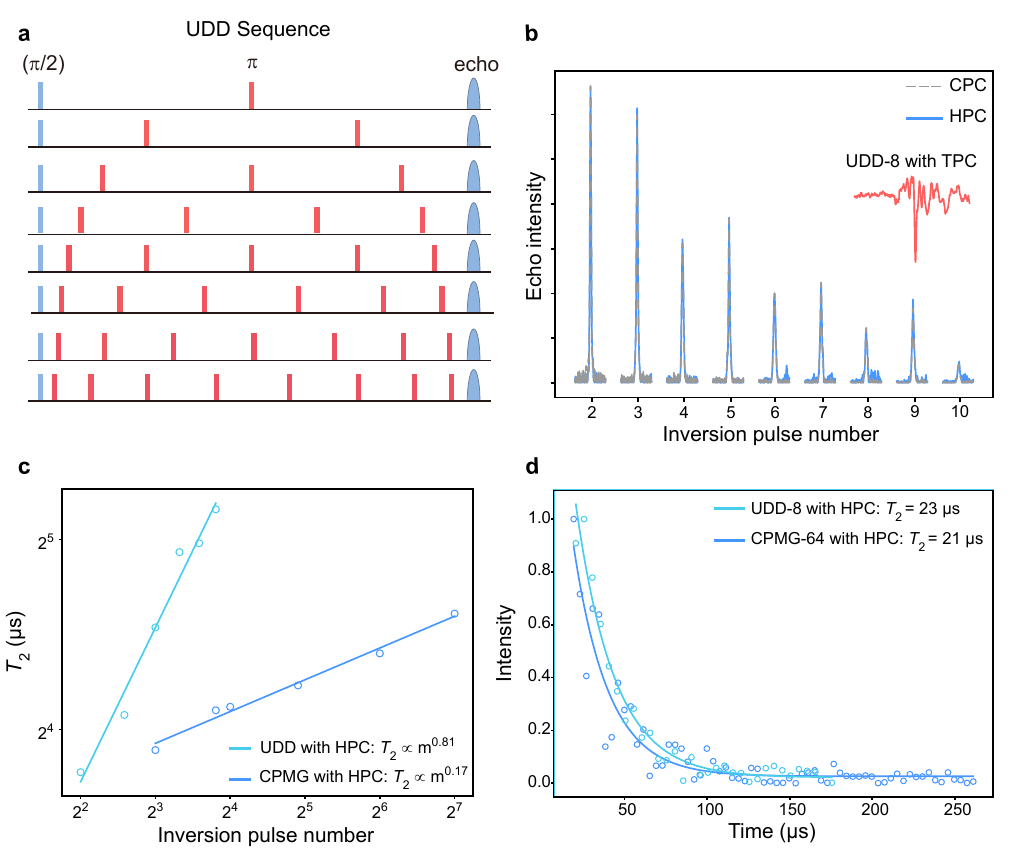}

		\caption{\textbf{Application of HPC for UDD on ensemble molecular qubits.}
			\textbf{a}, Schematic illustrations of UDD-1 to UDD-8. \textbf{b}, Comparison of echo intensities collected with HPC and CPC. Inset: Echo signals observed for UDD-8 with TPC. \textbf{c}, Dependencies of $T_2$ on the number of inversion pulses for UDD and CPMG with HPC. Both axes are in log scale. \textbf{d}, Comparison of echo decay curves under UDD-8 and CPMG-64 with HPC. Data were collected on ensemble Cu$^{2+}$-based molecular qubits at 8 K.}
		\label{fig:3}
	\end{figure}

\subsection*{Quantum error mitigation for single qubits}
To quantify the performance of phase cycling on single-qubit platforms, we introduce the effective state fidelity ($F^{eff}$). Because phase cycling averages the outputs of multiple quantum circuits prone to control and decoherence errors, the resulting state is fundamentally a mixed state. We therefore normalize the averaged state to its pure-state equivalent $\rho$, which effectively decouples the control-error mitigation performance from the irreversible decoherence. This normalized state $\rho$ is then compared to the ideal target state $\sigma$ using the standard fidelity metric:
\begin{equation}
	F^{eff}=(Tr \sqrt{\sqrt{\rho} \sigma\sqrt{\rho}})^2.
\end{equation}

We experimentally benchmarked TPC and HPC on single trapped ion and superconducting qubits with CP-$m$ and UDD-$m$ sequences that lack inherent robustness. The final state of each quantum circuit was characterized by quantum state tomography. For CP-$m$ with 2 -- 16 inversion pulses exerted on single trapped $^{40}$Ca$^+$ ion qubit, HPC retains $F^{eff}$ above 99.5$\%$ (Fig. 4c). In sharp contrast, the fidelity quickly declines to 58.3$\%$ for CP-16 with TPC. This trend was reproduced by UDD-$m$, where $F^{eff}$ is 99.5$\%$ on UDD-64 with HPC yet it plummets to 15.0$\%$ with TPC (Fig. 4d). Single superconducting transmon qubit exhibits similar behaviors, with TPC significantly deteriorating the effective state fidelity while HPC maintains remarkable robustness under relatively large systematic error (Fig. S7 and S8).

 The key advantage of HPC lies in its ability to separate the qubit's population and coherence, mimicking an ideal inversion pulse. It is evidenced by negligible z-component in the effective state (Fig. 4a), which is in stark contrast to the salient z-component observed in the final state of each individual quantum circuit (Fig. 4b and Fig. S8).    Together with the pathway-selection mechanism, the recovery of effective state fidelity under HPC strongly suggests that coherent control errors dominate the observed loss of effective state fidelity in the trapped-ion measurements. Small systematic pulse errors can accumulate coherently over a long sequence even when the single-pulse gate fidelity is relatively high. Rotation-angle error provides a plausible mechanism for this accumulation, without uniquely identifying the microscopic error source in the experiment (Supplementary Information). These results demonstrate the suppression of control-induced distortions in the measured signal. As dynamical decoupling is routinely used to protect quantum information of idle qubits during quantum algorithms\cite{ref3}, its integration with scalable phase cycling may improve the performance of quantum computing.

\begin{figure}[!htbp]
		\centering
		\includegraphics[width=\textwidth]{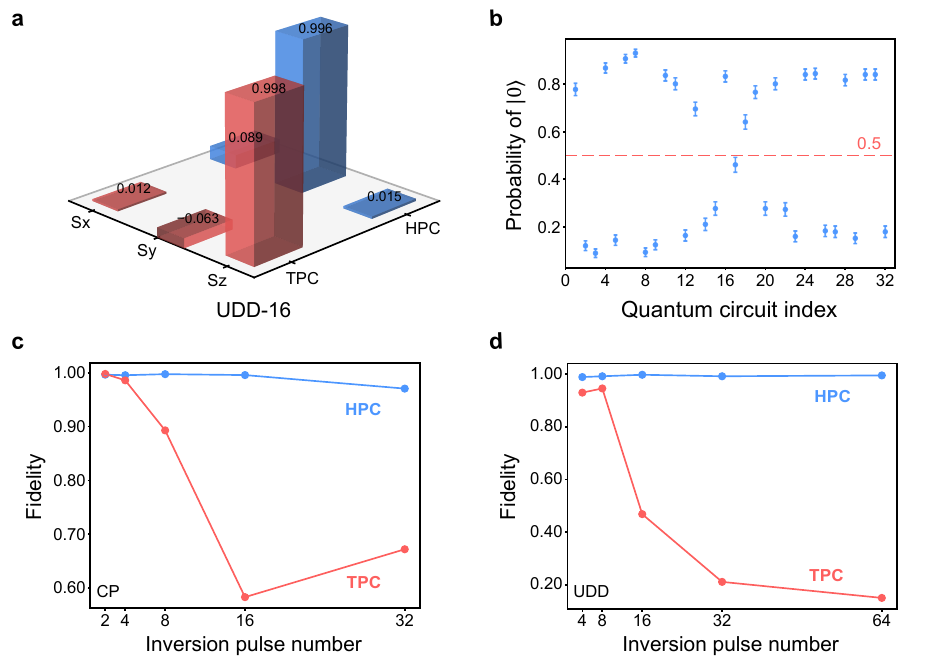}

		\caption{\textbf{HPC and effective state-fidelity tests on single trapped $^{\textbf{40}}$Ca$^+$ ion qubit.}
			\textbf{a}, Quantum state tomography results of effective final states. \textbf{b}, Probability of being at $\lvert 0 \rangle$ state for the final state of each quantum circuit. The average of all states is 0.508, indicating nearly-zero z-component for the effective final state. \textbf{c,d}, Effective state fidelity under \textbf{c} CP and \textbf{d} UDD with HPC and TPC.}
		\label{fig:4}
	\end{figure}

\section*{Discussion}

By leveraging the phase degree of freedom in quantum circuits and the abelian group structure, we designed Hadamard phase cycling as the first practical non-Markovian quantum error mitigation for inversion-pulse-based dynamical decoupling. This method balances effectiveness with scalability, enabling accurate measurement of $T_2$ and reliable implementation of dynamical decoupling with linear scaling for multiple leading qubit platforms. This extends phase cycling from complete pathway selection at modest sequence depths in magnetic resonance to a scalable quantum-information primitive for approximate pathway selection in deep sequences on NISQ-era devices.

 Our results demonstrate that residual control errors can introduce coherence-population mixing and prolong measured echo decays, particularly when $T_1\gg T_2$. This mechanism provides grounds to question whether coherence benchmarks obtained under comparable conditions reflect only the target transverse-coherence dynamics when population-mediated contributions have not been excluded. More generally, the interplay between control errors and platform-specific decoherence mechanisms makes it difficult to define a universal boundary for dynamical decoupling sequences without error mitigation. We therefore recommend directly comparing TPC and HPC measurements under matched conditions. HPC can strongly suppress population-mediated signal contributions, providing an experimental reference for assessing their influence on the measured decay. For applications concerned only with the final state rather than the evolution history, the benefit of HPC should instead be evaluated using an appropriate output-state or algorithm-level benchmark, such as quantum tomography.

Ultimately, these insights shift the focus of $T_2$ benchmarking toward task-specificity and explicit selection of the target coherence pathways. Furthermore, our study demonstrates that the utility of quantum error mitigation and suppression depends on a strategic balance of complexity and effectiveness. This integrated paradigm ensures that quantum control remains both scalable and physically grounded, offering new avenues for navigating the NISQ era and beyond.

\section*{Methods}
\subsection*{Qubit platforms}
Experiments on trapped ion qubits were conducted using a $^{40}$Ca$^+$ ion with Paul trap based on a CIQTEK ION I quantum processor, whose single gate fidelity was above 99.9$\%$. Experiments on superconducting qubits were conducted using a transmon qubit of the Starmon-7 processor, accessed via the Quantum Inspire platform, whose single gate fidelity was well above 99$\%$. X-band pulse electron paramagnetic resonance (EPR) spectroscopic characterization of molecular qubits was performed at 8 K on a CIQTEK EPR100 spectrometer. The Cu$^{2+}$-based molecular qubit, (PPh$_{4}$)$_{2}$[Cu(mnt)$_{2}$] (PPh$_4^+$ = tetraphenylphosphonium; mnt$^{2-}$ = 1,2-maleonirtrileditholate), was synthesized following literature procedure\cite{ref37}, and it was diluted in a diamagnetic isostructural host (PPh$_{4}$)$_{2}$[Ni(mnt)$_{2}$] with a concentration of 0.3$\%$\cite{ref20}. This molecule exhibits $T_1$ = 37.03 ms and $T_2^{Hahn}$ = $6.54~\mathrm{\mu s}$ revealed by inversion recovery and Hahn echo decay experiments, respectively (Fig. S3). The radical-based molecular qubit, COF-5, was synthesized following literature procedure\cite{ref40}. NV Centers used in this work have an average depth of 30 nm from the surface of diamond. Optical detected magnetic resonance experiments were performed with a homebuilt standard setup under an external magnetic field of approximately 315 G, with microwave phase errors less than 1 degree and frequency detuning less than 1.5 MHz. At room temperature, this sample exhibits $T_1$ = $1.93~\mathrm{ms}$ and $T_2^{Hahn}$ = $9.29~\mathrm{\mu s}$ (Fig. 2e and Fig. S6).
\subsection*{Dynamical decoupling sequences}
CP sequences were implemented as $\pi_x/2-[\tau-\pi_x-\tau]_n-echo$. CPMG sequences were implemented as $\pi_x/2-[\tau-\pi_y-2\tau-\pi_y-\tau]_n-echo$, where the phase of inversion pulse is shifted by $\pi/2$ from the phase of the first $\pi$/2 pulse. The first inversion pulse's error within each block can be compensated by the following inversion pulse, giving rise to its inherent robustness. The implemented UDD-$m$ involves $m$ nonperiodic inversion pulses following a $\pi_x/2$ pulse, where all pulses have the same phase and the $n^{th}$ inversion pulse is applied at\cite{ref19,ref42}:
\begin{equation}
	t_n=tsin^2\frac{n\pi}{2(m+1)},n=1,\dots,m.
\end{equation}

\subsection*{Two-step phase cycling}
TPC involves two quantum circuits that differ only in phase of the first $\pi/2$ pulse; one uses a $\pi_x/2$ pulse and the other uses $\pi_{-x}/2$ pulse. Subtracting the results of these two circuits effectively cancels baseline drift. For experiments on single trapped ion, superconducting qubits, and NV centers in diamond, an extra $\pi/2$ pulse is applied before measuring population distribution. In this case, phase cycling the final $\pi/2$ pulse also suffices.

\subsection*{Complete phase cycling}
For IDD-$m$, the aforementioned CPC involves $2^m$ quantum circuits encompassing all possible phase configurations of inversion pulses\cite{ref19,ref30}. Practically, combination of CPC and TPC is required, where the phase of the first inversion pulse can be fixed so that the complexity does not change.

\subsection*{Hadamard phase cycling}
Sylvester's method was used to construct Hadamard matrices\cite{ref39}:
\begin{equation}
	H_{2^{n}}=
	\begin{pmatrix}
		H_{2^{n-1}} & H_{2^{n-1}} \\
		H_{2^{n-1}} & -H_{2^{n-1}}
	\end{pmatrix}, H_2=
	\begin{pmatrix}
		1 & 1 \\
		1 & -1
	\end{pmatrix}.
\end{equation}
where integer $n\ge2$. For IDD-$m$, one needs to construct the smallest Sylvester's Hadamard matrix with dimension $2^n\ge m$ and drop the redundant columns. This property gives HPC's complexity between $4m$ and $8m-8$. For simplicity, the orthogonality ratio of HPC was derived analytically for $m=2^n$ by analyzing HPC's algebraic structure. All combinations of odd-numbered columns from $H'$ satisfy Hadamard product orthogonality (supplementary information), and some combinations of even-numbered columns from $H'$ do not. For an even number $2\le2k<2^n$, given $C_{2^n-1}^{2k}$ as the number of all possible combinations of choosing $2k$ columns from $2^n-1$ columns in $H'$, the ratio of combinations that do not satisfy Hadamard product orthogonality is given by
\begin{equation}
	p(2k)=\frac{1}{2^n}+(-1)^{k}\times\frac{2^n-1}{2^n}\times\frac{C_{2^{(n-1)}-1}^k}{C_{2^{n}-1}^{2k}}.
\end{equation}
\subsection*{Effective state fidelity}
Define $\rho^i$ as the final state of the $i^{th}$ quantum circuit, and define phase-cycled expectation values as
\begin{equation}
	v_{x,y,z}=\sum_{i}V_s^iTr(\rho^{i}S_{x,y,z}),
\end{equation}
where $V_s^i$ is the $i^{th}$ element of the sign vector; $Tr(\rho^i S_x )$ was obtained by adding a $\pi_{(-y)}/2$ pulse before measurement, and $Tr(\rho^i S_y )$ was obtained by adding a $\pi_{x}/2$ pulse. The normalized $v_{x,y,z}$ were used to construct the effective state
\begin{equation}
	\rho=\frac{1}{2}(S_{0}
	+ \left \langle v_x \right \rangle S_{x}
	+ \left \langle v_y \right \rangle S_{y}
	+ \left \langle v_z \right \rangle S_{z}).
\end{equation}
where $S_0$ is two-dimensional identity matrix.

\subsection*{Data availability}
The data supporting our findings are included in the article and supplementary files. Additional data generated during the study are available from the corresponding author upon request.

\section*{Acknowledgments}
This work was supported by the National Natural Science Foundation of China (grant No. 22273078, No. 92576117, and No. T2125011), the Hangzhou Municipal Funding Team of Innovation (TD2022004), and the Chinese Academy of Sciences (grant No. YSBR-068). W.N. and L.S. acknowledge CIQTEK Co., Ltd., Quantum Inspire, and the Instrumentation and Service Center for Molecular Sciences at Westlake University for facility support and technical assistance. W.N. and L.S. acknowledge Dr. Zhifu Shi for providing the Cu$^2+$-based molecular qubit, Qian Cao for preliminary theoretical studies, Junfeng Wu, Dr. Zhichao Liu, and Dr. Mengxiang Zhang for assistance with experiments on trapped ion qubits, Marios Samiotis from Dicarlo laboratory for assistance with experiments on superconducting transmon qubits, and Dr. Maria Grazia Concilio and Prof. Xueqian Kong for preliminary simulations. W.N. acknowledges Dr. Aimei Zhou and Zhiyang Wang for assistance with pulse EPR spectroscopy, Jiaqi Han, Prof. Jian Li, and Prof. Huayi Chen for helpful discussions, and Yingchao Wang and Xiaoqing Yuan for assistance with artistic design. L.S. acknowledges Prof. Tijana Rajh, Dr. Tianxing Zheng, Prof. Dafei Jin, Prof. Xinhao Li, Dr. Ralph T. Weber, Dr. Yutian Wen and Prof. Lorenza Viola for helpful discussions.
\section*{Author contributions}
L.S. and W.N. conceived the idea, designed experiments, and oversaw the project. W.N. conducted theoretical analysis, proposed HPC and its benchmark, and performed experiments on trapped ion, superconducting, and molecular qubits. Z.L., G.Q., and F.S. performed experiments on NV centers in diamond, and W.N. assisted with these experiments. A.E. performed experiments on nuclear spins with solid-state NMR. Z.S. synthesized radical-based molecular qubits. J.D. assisted with theoretical analysis. W.N. and L.S. co-wrote the manuscript. All authors have given approval to the final version of the manuscript.
\section*{Competing interests}
There are no competing interests to declare.

\endgroup
\clearpage
\begingroup

\renewcommand{\thefigure}{S\arabic{figure}}
\renewcommand{\thetable}{S\arabic{table}}
\renewcommand{\theequation}{S\arabic{equation}}
\renewcommand{\thepage}{S\arabic{page}}
\setcounter{figure}{0}
\setcounter{table}{0}
\setcounter{equation}{0}

\setcounter{page}{1}
\renewcommand{\theHfigure}{S\arabic{figure}}
\renewcommand{\theHtable}{S\arabic{table}}
\renewcommand{\theHequation}{S\arabic{equation}}

\begin{center}
	\section*{Supplementary Information for Scalable Quantum Error Mitigation for Dynamical Decoupling via Hadamard phase cycling}

	Weibin~Ni,
	Zhijie~Li,
	Guanyu~Qu,
	Asif~Equbal,
	Zhecheng~Sun,
	Jiale~Dai,
	Fazhan~Shi,
	Lei~Sun$^\ast$\\
	\small$^\ast$Corresponding author. Email: sunlei@westlake.edu.cn\\
\end{center}

\subsection*{Materials and Methods}
\subsubsection*{Ensemble electron spin qubits in molecules}
\textbf{(PPh$_4$)$_2$[Cu(mnt)]$_2$}:
The molecular electron spin qubit, (PPh$_4$)$_2$[Cu(mnt)$_2$], was synthesized following literature procedure\cite{si_ref37}. It was diluted in a diamagnetic isostructural host (PPh$_4$)$_2$Ni(mnt)$_2$ with a concentration of $0.3\%$ via co-crystallization. This molecule exhibits $T_1=37.03~\mathrm{ms}$ and $T_2^{Hahn}=6.54~\mathrm{\mu s}$ revealed by inversion recovery and Hahn echo decay experiments, respectively. Pulse electron paramagnetic resonance (EPR) spectroscopic characterization of this molecule was performed with resonant frequency of 9.6623 GHz under $3340~\mathrm{G}$ magnetic field at $8~\mathrm{K}$.
\\
\textbf{COF-5}:
COF-5 was synthesized from 2,3,6,7,10,11-hexahydroxytriphenylene and benzene-1,4-diboronic acid based on the literature procedure\cite{si_ref40}. It hosts semiquinone-like radical qubits derived from partially oxidation of the former linker. It exhibits decent spin dynamic properties at room temperature with $T_1=14.13~\mathrm{ms}$ and $T_2=4.00~\mathrm{\mu s}$. Pulse EPR spectroscopic characterization of this molecule was performed with resonant frequency of $9.6776~\mathrm{GHz}$ under $3350.5~\mathrm{G}$ magnetic field at $80~\mathrm{K}$.
\subsubsection*{Solid-State NMR (SSNMR)}
$^{13}$C nuclear spin in a labeled glycine was used as qubit for measurements, with $T_1=12$ s and $T_2^{Hann}=8573~\mathrm{\mu s}$. The sample was purchased from cambridge isotopes, and packed into a 1.3 mm zirconia rotor for magic-angle spinning (MAS) experiments. All SSNMR measurements were performed on a 14.4 T (600 MHz), wide-probe, spectrometer, using a 1.3 mm HXY fast MAS probe. The probe was shimmed using adamantane. The X-channel was tuned to $^{13}$C at a frequency of approximately 150 MHz. The sample was spun at the magic angle with a spin rate of 60 kHz to attenuate coherence dephasing effect from the  first order chemical shift and the dipolar coupling anisotropy.

To prevent interference between the radiofrequency (rf) pulses and the spatial rotation of the sample, all refocusing pulses were synchronized with the rotor spinning. For CPMG experiments, the $T_2$ dephasing profile was obtained by varying the number of $\pi$ pulses, with each pulse spaced by $n\tau_r$, where $n$ was fixed at 200 and $\tau_r$ corresponded to the duration of a single rotation cycle. Following the protocol described in the main text, HPC was applied to the $^{13}$C channel to mitigate control errors and isolate pure coherence pathways. Measurements were also performed using standard two-step phase cycling to benchmark the effectiveness of HPC in preventing $T_2$ overestimation due to coherence-population mixing.

Similar to the EPR experiments on molecular spins, nuclear spin qubits in the solid state are susceptible to control-induced artifacts. By using a high MAS rate of 60 kHz and rotor-synchronized pulses, we ensure that the $T_2$ measurements reflect the intrinsic decoherence of the $^{13}$C spins under the applied dynamical decoupling sequences.

\subsubsection*{Ensemble electron spin qubits in diamond NV centers}
The ensemble NV centers were created by implanting $^{15}$N$^+$ ions at an energy of $10~\mathrm{keV}$, and a dose of $5\times10^{12}~~\mathrm{cm^{-2}}$. The mean depth of implanted ions is approximately $30~\mathrm{nm}$. At this depth, the NV centers are sufficiently shielded from surface-induced noise and exhibit spin dynamic properties similar to those of deep NV centers. Prior to experiments, the diamond sample was chemically cleaned in a hot tri-acid mixture (H$_2$SO$_4$ : HNO$_3$ : HClO$_4$ = 1 : 1 : 1 by volume) at $453\ \mathrm{K}$. At room temperature, this sample exhibits $T_1=1.93~\mathrm{ms}$ and $T_2^{Hahn}=9.29~\mathrm{\mu s}$.

\subsubsection*{Trapped ion qubits}
Experiments on trapped ion qubits were conducted on the CIQTEK's Ion I Quantum Computer, where each $^{40}$Ca$^+$ ion was confined in a segmented-blade Paul trap. The qubit was manipulated on the 729 nm optical transition at room temperature, with single-qubit gate fidelity being approximately 0.999. The relaxation time was measured to be $T_1=1.0~\mathrm{s}$ using the inversion recovery technique. The coherence time was $T_2^{Hahn}=$ $976~\mathrm{\mu s}$ obtained from Hahn echo decay experiments.

\subsubsection*{Supercoducting qubits}
Experiments on superconducting transmon qubits were conducted on the Quantum Inspire's Starmon-7 Superconducting Quantum Processor, which consists of seven transmon qubits and is available on the cloud. Each transmon qubit was controlled via dedicated microwave and flux-bias lines for single-qubit and two-qubit gates, and was read out via a dispersively coupled resonator. Each resonator incorporated a Purcell filter and was frequency-multiplexed onto one of two shared feedlines: one for qubits Q$_0$ and Q$_2$, and another for the rest. More detailed information about hardwares of Starmon-7 can be found in
\url{ https://github.com/DiCarloLab-Delft/QuantumInspireUtilities/blob/main/Starmon7_FactSheet.pdf}. During our experiments, the single gate fidelity was well above $99\%$. Detailed information about the processors performance can be found on \url{https://monitoring.qutech.support/public-dashboards/7171f0e3cfc44995a97dff9001c4d7d1?orgId=1&from=now-7d&to=now}.

\subsubsection*{Pulse EPR spectroscopy}
X-band (around 9.6 GHz) pulse EPR spectroscopy was performed on a CIQTEK EPR100 spectrometer. The length of $\pi$ pulse was fixed to be $32~\mathrm{ns}$ for all experiments, determined by nutation experiments.
\\
\textbf{Inversion recovery measurements}:
The spin-lattice relaxation time ($T_1$) was measured using an inversion recovery sequence $\pi-T-\pi/2-\tau-\pi-\tau-echo$. For all experiments, a four-step phase cycling  was employed with pulse phases of $(+x, -x, +x)$, $(+x, +x, +x)$, $(-x, -x, +x)$, and $(-x, +x, +x)$. The resulting echo decay was measured by varying the interval $T$, and the inversion recovery curve was fitted with a stretched exponential decay function for (PPh$_4$)$_2$[Cu(mnt)$_2$] or a double-exponential decay function for COF-5 to extract $T_1$.
\\
\textbf{Hahn echo decay measurements}:
$T_2^{Hahn}$ was characterized using Hahn echo pulse sequence $(\pi/2-\tau-\pi-\tau-echo).$ Two-step phase cycling was used to cancel baseline drift. The Hahn echo decay was measured by varying the interval $\tau$, and the echo decay curve was fitted with a stretched exponential decay function for (PPh$_4$)$_2$[Cu(mnt)$_2$] or a mono-exponential decay function for COF-5 to extract $T_2$.
\\
\textbf{CPMG measurements}:
Typical CPMG sequence can be represented as $(\pi_x/2-[\tau-\pi_y-2\tau-\pi_y-\tau]_n-echo)$. We conducted two types of CPMG experiments. In the first type, while fixing pulse intervals ($\tau=120,150,400~\mathrm{ns}$
for NV centers, (PPh$_4$)$_2$[Cu(mnt)$_2$], and COF-5 respectively) and adding pulse number, the echo at time $\tau$ after each pulse was collected. In the second type, while fixing the pulse number and increasing the pulse interval, we monitored the decay of the echo at time $\tau$ after the last pulse, which can be fitted by a mono-exponential decay function to extract $T_2$. For 5 of the 14 Cu$^{2+}$ CPMG decays in Table~\mbox{\ref{tab:mono-fit-residuals}}, allowing the stretch exponent to vary yields similarly good fits for multiple parameter sets with different $T_2$ values, making $T_2$ difficult to determine reliably. We therefore use the mono-exponential model to define a common apparent decay time for the HPC--TPC comparison.

For the CPMG and UDD decays in Table~\mbox{\ref{tab:mono-fit-residuals}}, normalized residuals were calculated as $r_i=[y_i-f(t_i)]/\max_j y_j$, where $y_i=|S(t_i)|$ is the absolute value of the measured signal and $f(t_i)$ is the mono-exponential fit. The table reports their root-mean-square (RMS) value, $\sqrt{N^{-1}\sum_{i=1}^{N}r_i^2}$, using all $N$ recorded points.
\\
\textbf{UDD measurements}:
For UDD-$m$, the inversion pulses were applied at time $t_n=tsin^2\frac{n\pi}{2(m+1)},n=1,2,\dots,m$. The desired echo appeared at time $t$. Similar to CPMG experiments, we also conducted two types of UDD-$m$ experiments. In the first type, while fixing $t=$ 5000 ns and adding pulse number, the desired echo after each pulse was collected. In the second type, while fixing the pulse number and increasing $t$, the decay curve of the desired echo after the last pulse was collected and fitted by a mono-exponential decay function to extract $T_2$. For consistency with the CPMG comparison, we use the same mono-exponential model to define an apparent decay time for UDD; this choice does not establish single-exponential microscopic relaxation.

\subsubsection*{Optical detected magnetic resonance spectroscopy}
Experiments for NV centers in diamond were performed with a homebuilt standard setup under a quantizing magnetic field of approximately 315 G, with microwave phase errors less than 1 degree and frequency detuning less than 1.5 MHz. The mono-exponential decay function was used for fitting to extract $T_1$ and $T_2$.

\subsubsection*{Effective state fidelity test}
\textbf{Quantum state tomography}:
As mentioned in supplementary text (see equation $(S40)$), $s_{x,y,z}^i$ are required for state reconstruction. To obtain these values, $\frac{\pi_{x,-y}}{2}$ pulses were applied before measurements for $s_{y,x}^i$. No extra pulse was needed to measure $s_{z}^i$.
\\
\textbf{Test on the trapped ion qubit}:
The tests involve CP-$m$ and UDD-$m$. CP-$m$ can be represented as $(\pi_x/2-(\tau-\pi_x-\tau)_m-echo)$. For CP-m experiments, the $\tau$ value was fixed to be $50~\mathrm{\mu s}$. For the UDD-$m$ sequences, the $t$ value was fixed to be $1200~\mathrm{\mu s}$. TPC's results were obtained from an average of 2048 shots for each phase cycling step; HPC's results were obtained from an average of 256 shots for each phase cycling step.
\\
\textbf{Test on the superconducting qubit}:
For CP-$m$ experiments, the $\tau$ value was fixed to be 250 ns. For the UDD-$m$ sequences, the $t$ value was fixed to be $100~\mathrm{\mu s}$. HPC results were obtained from an average of 1024 shots for each phase cycling step. TPC results were obtained similarly. Considering that TPC involves only two phase cycling steps, we intentionally added redundant experiments to align with HPC's number of shots.
\\
\textbf{Results of the superconducting qubit}:
Regarding UDD-$m$, HPC preserves $F^{eff}$ above $99.3\%$ up to 16 inversion pulses, whereas TPC gives rise to significantly smaller values, with $F^{eff}$ dropping to $97.3\%$ for UDD-16 and $53.5\%$ for UDD-64. We further deliberately introduced systematic error to each pulse. For CP-m with $\pi/28$ systematic error, HPC retains $F^{eff}$ above $99.3\%$ for m = 2-128, whereas TPC causes an oscillation of $F^{eff}$, which drops to 0 for m being 32 and 96. Therefore, HPC maintains remarkable robustness even under relatively large systematic error.

\subsection*{Supplementary Text}
\subsubsection*{Problem setup}
We consider a two-level quantum system (qubit) subject to pure dephasing errors. Without loss of generality, one can imagine it as an ensemble of electron spins ($S = \frac{1}{2}$) under a static and strong quantizing magnetic field $B_{0}$ along the $z$-axis, with pure dephasing errors described by the on-site disorder:
\begin{equation}
	H=(\omega+\delta \omega_{i}(t))\hat{S_{z}},
\end{equation}
where $\delta \omega_{i}(t)\ll \omega$ and $\omega=g_{e}\mu_{B}\hbar^{-1}B_{0}$ is unperturbed Zeeman splitting with $g_{e}$, $\mu_{B}$, $\hbar$, $\hat{S}_z$ representing the $g$-factor of the electron spin, Bohr magneton, reduced Planck constant, and a Pauli-$z$ matrix, respectively. We further assume that all spin-spin interaction terms are negligible such that $\hat{S}_z$ is the only non-trivial term.

Due to the negligible spin-spin interactions, we focus on a single spin in the rotating frame for our analysis, where the first term $\omega\hat{S_{z}}$ can be omitted. The system's state is fully described by the $2 \times 2$ density matrix ${\rho}$. We assume the initial state to be:
\begin{equation}
	\rho(0)=
	\begin{pmatrix}
		1 & 0 \\
		0 & 0
	\end{pmatrix}.
\end{equation}
In general, the qubit's density matrix reads:
\begin{equation}
	\rho=
	\begin{pmatrix}
		P_{0} & C_{+} \\
		C_{-} & P_{1}
	\end{pmatrix},
\end{equation}
where
\begin{itemize}
	\item $P_{0}$ and $P_{1}$ represent populations of states $\left|0\right\rangle$ and $\left|1\right\rangle$, respectively, with $P_0=1-P_1$;
	\item $C_{+}$ and $C_{-}$ are complex conjugates ($C_{-}=C_{+}^*$) encoding the coherence between $\left|0\right\rangle$ and $\left|1\right\rangle$.
\end{itemize}
To understand why coherence-population mixing matters, we add longitudinal relaxation characterized by $T_1$ and transverse relaxation characterized by $T_2$:
\begin{equation}
	\rho(t)=
	\begin{pmatrix}
		1-P_1e^{-t/T_1} & C_+e^{-i\delta\omega t}e^{-t/T_2} \\
		C_-e^{i\delta\omega t}e^{-t/T_2} & P_1e^{-t/T_1}
	\end{pmatrix}\\.
\end{equation}
Equivalently, the density matrix can be represented as:
\begin{equation}
	\rho(t)=\frac{1}{2}\left [ \hat{S}_0+(1-2P_1e^{-t/T_1})\hat{S}_z+C_+e^{-i\delta \omega t-t/T_2}\hat{S}_++C_-e^{i\delta \omega t-t/T_2}\hat{S}_- \right ] .
\end{equation}

An ideal inversion pulse would exchange populations ($P_0 \leftrightarrow P_1$) and swap coherences ($C_{+} \leftrightarrow C_{-}$) with a phase factor. However, imperfect (noisy) inversion pulses introduce coherence-population mixing, generating spurious signals like stimulated echoes that adversely affect coherence measurements.

The density matrix can be represented by a linear combination of a complete basis. The most commonly used basis set is $\left \{ \hat{S}_0,\hat{S}_x,\hat{S}_y,\hat{S}_z\right \}$, where $\hat{S}_0$ is the two-dimensional identity matrix and $\hat{S}_{x,y,z}$ are Pauli matrices. Here, we choose $\left \{ \hat{S}_0,\hat{S}_+,\hat{S}_-,\hat{S}_z\right \}$ because these terms explicitly represent coherence and population. It allows straightforward calculation of the free evolution with pure dephasing:

\begin{equation}
	\begin{aligned}
		\hat{S}_0(t) &=  \hat{S}_0(0), \\
		\hat{S}_z(t) &=  \hat{S}_z(0), \\
		\hat{S}_+(t) &=  \hat{S}_+(0)e^{-i\delta \omega t}, \\
		\hat{S}_-(t) &=  \hat{S}_-(0)e^{+i\delta \omega t},
	\end{aligned}
\end{equation}
where $\hat{S}_{\pm}=\frac{1}{2}(\hat{S}_x\pm i\hat{S}_y)$. We can then define coherence orders $p=\pm1,0$:
\begin{equation}
	\hat{S}_p(t) =  \hat{S}_p(0)e^{-ip\delta \omega t}.
\end{equation}
Note that
\begin{equation}
	\hat{S}_{+}=
	\begin{pmatrix}
		0 & 1 \\
		0 & 0
	\end{pmatrix}, \hat{S}_{-}=
	\begin{pmatrix}
		0 & 0 \\
		1 & 0
	\end{pmatrix},\frac{\hat{S}_{0}+\hat{S}_{z}}{2}=
	\begin{pmatrix}
		1 & 0 \\
		0 & 0
	\end{pmatrix},\frac{\hat{S}_{0}-\hat{S}_{z}}{2}=
	\begin{pmatrix}
		0 & 0 \\
		0 & 1
	\end{pmatrix}.
\end{equation}
Evidently, $p=\pm1$ represent coherence, and $p=0$ represents population (see equation $(S3)$). For simplicity, we drop the identity term in the following analysis.

\subsubsection*{Hahn echo}
The above basis enables calculation of the most well-known spin echo, namely Hahn echo, generated by a $\pi/2$ pulse followed by an inversion pulse ($\pi$ pulse):
\begin{equation}
	\hat{S}_{z}\xrightarrow{(\frac{\pi}{2})_{y}}\frac{\hat{S}_{+}+\hat{S}_{-}}{2}\xrightarrow{\tau}\frac{\hat{S}_{+}e^{-i\delta\omega\tau}+\hat{S}_{-}e^{i\delta\omega\tau}}{2}\xrightarrow{(\pi)_{x}}\frac{\hat{S}_{-}e^{-i\delta\omega\tau}+\hat{S}_{+}e^{i\delta\omega\tau}}{2}
\end{equation}

After another free evolution time $\tau$, $\frac{\hat{S}_{+}+\hat{S}_{-}}{2}$ reappears, resulting in a Hahn echo. This echo reflects decoherence as it only involves $p=+1$ and $p=-1$ after the $\pi/2$ pulse. Note that $\delta \omega$ is assumed to be constant within time $2\tau$, and we apply this assumption throughout the following analysis for simplicity. We can also track the coherence order transfer during this process. There are two symmetric pathways: $0\to+1\to-1$ and $0\to-1\to+1$. Due to symmetry, we only need to consider one of them as $\hat{S}_{+}$ and $\hat{S}_{-}$ always appear in pairs with equal magnitude.

\subsubsection*{Conclusion 1: Echo formation condition}
The coherence order of the initial state is $p=0$ according to equation $(S2)$ and equation $(S7)$. We can then represent the coherence transfer pathways by two series of coherence orders:
\begin{equation}
	(0,p_{1},p_{2},...,p_{n},-1),(0,-p_{1},-p_{2},...,-p_{n},+1).
\end{equation}
They are symmetric and essentially summarize the same dynamics. The echo formation condition requires the qubit to spend equal time in $\hat{S}_{+}$ and $\hat{S}_{-}$ because all dephasing will be refocused by the inversion pulse according to equation $(S6)$.

For free evolution times $(\tau_{1},\tau_{2},...,\tau_{n})$ corresponding to each coherence order, if the final coherence order is $-1$, the echo occurs at time $t$ after the last pulse when the following condition is satisfied:
\begin{equation}
	t=\sum_{i=1}^{n}p_{i}\tau_{i}>0.
\end{equation}
This means that the qubit spends more time on $p=+1$ than $p=-1$, so that it needs some extra time on $p=-1$ to form an echo. If the final coherence order is $+1$, then $t$ has to be smaller than $0$. Assume that the qubit has evolved according to the first coherence transfer pathway in equation $(S10)$. It gains a phase of $e^{-i\delta\omega\sum_{i=1}^{n} p_{i}\tau_{i}}=e^{-i\delta\omega t}$. After free evolution time $t$ on $p=-1$, this extra phase disappears according to equation $(S6)$, generating an echo.

\subsubsection*{Stimulated echo}
The classic pulse sequence generating a stimulated echo is: $(\frac{\pi}{2}-\tau-\frac{\pi}{2}-T-\frac{\pi}{2})$. While there exist numerous coherence transfer pathways, Conclusion 1 immediately reveals that a stimulated echo will appear at time $\tau$ after the last pulse, corresponding to the pathway $(0,+1,0,-1)$ or $(0,-1,0,+1)$. Pathways that do not form echoes (anti-echoes) are typically much weaker due to decoherence. The formation of stimulated echo involves coherence-population mixing after each pulse, so the dynamics of this echo reflects both decoherence and longitudinal relaxation.

\subsubsection*{Conclusion 2: Phase shift of signal}
Note the $i^{th}$ pulse induces a change in the coherence order $\Delta p_{i}$, and a phase change $\Delta\phi_{i}$ in the $i^{th}$ pulse causes a phase shift in the final signal. Accordingly, the total phase shift $\Delta\Phi$ in the final signal is:
\begin{equation}
	\Delta\Phi=\sum_{i}\Delta p_{i}\Delta\phi_{i}.
\end{equation}
In this context, all phases are defined as the angle with respect to $+x$ axis. For example, a pulse along $+y$ has a $\frac{\pi}{2}$ phase, and the corresponding single qubit rotation is along $+y$ axis. An echo with phase $+y$ means the echo appears at $+y$ axis. In our setup, $\Delta p\in\left \{  0,\pm1,\pm2\right \}$ and $\Delta\phi\in  \left\{ 0,\pi\right \}$. This implies that changing a pulse's phase introduces a $\pi$ phase shift to the final signal only when $\Delta p=\pm1$. The coherence transfer pathways can thus be classified by the number of $\Delta p=\pm1$ transitions. This classification aligns perfectly with our goal of distinguishing population and coherence. For $k$ phase-flipped pulses:
\begin{itemize}
	\item Even $k$: No change in the final signal ($\Delta\Phi=0$)
	\item Odd $k$: Final signal gains a minus sign ($\pi$ phase shift, i.e. $\Delta\Phi=\pi$)
\end{itemize}
These properties form the foundation of phase cycling design for inversion-pulse-based dynamical decoupling (IDD).

\subsubsection*{Phase cycling}
Phase cycling enables selection of desired coherence transfer pathways and corresponding echoes. A noisy pulse could cause all possible coherence order transfers, and an $N$-pulse sequence can possibly generate $\frac{1}{2}(3^{N-1}-1)$ echoes, making the phase cycling of dynamical decoupling sequences complicated.

Take CPMG-$m$ sequence as an example, which includes a $\pi/2$ pulse followed by $m$ spin-locking inversion pulses with equal intermittent free evolution times. We first consider the CPMG-2 sequence with fixed final coherence order $p=-1$ to show the principle of phase cycling. There are 4 coherence transfer pathways that could generate echoes after the last pulse as summarized in Table S1.

The desired pathway $(0,-1,+1,-1)$ representing qubit coherence overlaps with $(0,+1,0,-1)$ (stimulated echo). Phase cycling resolves this issue: the desired echo is an one-pulse echo while the stimulated echo is a three-pulse echo, distinguished by the number of $\Delta p=\pm1$ transitions within their pathways (Table S2). Later we will see such a classification allows us to analyze $\frac{1}{2}(3^{N-1}-1)$ echoes in an elegant way. In this case, by performing two experiments, where one of them flips either inversion pulse with respect to the other experiment, and summing the final signals, the stimulated echo cancels out while the desired echo doubles (see Table S3 for a more complex phase cycling that involve four experiments).

\subsubsection*{Phase cycling problem in CPMG}
Recall that a coherence transfer pathway can be represented as:
$$(0,p_{1},p_{2},...,p_{m+1},-1).$$ The key to designing the phase cycling for CPMG-$m$ is to exclude $p_i=0$ within the coherence transfer pathway because we should probe pure decoherence processes ($p_{i}=\pm1$) without involving the longitudinal relaxation. Our objective then reduces to eliminating $\Delta p=\pm1$ transitions, which are generated by imperfect inversion pulses, except for the first $\pi/2$ pulse that prepares the coherence.

Consider the stimulated echo in the example of phase cycling in Table S3. Denote the original pulse phases (the first row) to be $\left [+1,+1,+1 \right ]$. These pulses produce the final signal with the original phase. Phase flipping of any pulse ($+1$ to $-1$, i.e. $\Delta\phi_i=\pi$) inverts the signal sign, producing a $\pi$ phase shift of the final signal. Critically, the phase of the final signal is determined by the product of all pulse phases. For instance:
\begin{itemize}
	\item $\left [+1,+1,-1\right ]$ and $\left [+1,-1,+1 \right ]$ yield the final signal with a $\pi$ phase shift;
	\item $\left [+1,-1,-1 \right ]$ gives the final signal with the original phase.
\end{itemize}
Denote the original and $\pi$ phase-shifted signals as $+1$ and $-1$, respectively. The sum of results of the four phase cycling steps is $+1+(-1)+(-1)+1=0$, confirming that the phase cycling cancels out the three-pulse echo, i.e. stimulated echo. In contrast, the desired echo is amplified because the phase of the first $\pi/2$ pulse is not flipped. Consequently, we can transform the design principle of phase cycling into an orthogonality condition that is concisely expressed as:
\begin{equation}
	\begin{pmatrix}
		+1\\
		+1\\
		+1\\
		+1
	\end{pmatrix}\circ \begin{pmatrix}
		+1\\
		+1\\
		-1\\
		-1
	\end{pmatrix}\circ \begin{pmatrix}
		+1\\
		-1\\
		+1\\
		-1
	\end{pmatrix}=\begin{pmatrix}
		+1\\
		-1\\
		-1\\
		+1
	\end{pmatrix}.
\end{equation}
Here, ``$\circ$''  denotes the Hadamard product, i.e. the element-wise product. The orthogonality requires the summation of all elements in the result vector to be zero. In this case, three-vector orthogonality translates to the cancellation of the three-pulse echo. As the phase of the desired echo is exclusively determined by the the phase of the first $\pi/2$ pulse, this orthogonality condition is applicable to all IDD sequences.

\subsubsection*{Basic definitions for phase cycling}
Define a phase cycling set:
\begin{equation}
	G \subseteq V_{N}, \quad G= \{ g_{1}, g_{2}, \dots, g_{n} \} \cup \{ E \}
\end{equation}
where $V_{N}$ is an $N$-dimensional vector space, each $g_{i}$ contains equal numbers of $+1$ and $-1$ elements, and $E$ is an all-one vector. The phase cycling matrix comprises elements from this set that are used as column vectors. Each row of this matrix represents phases of all pulses within the sequence in one phase cycling step, and each column represents the phase configuration of one pulse throughout all phase cycling steps.

Define Hadamard product orthogonality (HPO) of $k$ column vectors in $G$ if they fulfill the following condition:
\begin{equation}
	g_{1}\circ g_{2}\circ \dots \circ g_{k} = g_{i} \neq E.
\end{equation}
As the sum of all elements in $g_i$ is equal to $0$, this condition leads to the cancellation of k-pulse echoes.

\subsubsection*{Complete phase cycling (CPC)}
\textbf{Construction}:
We now prove that CPC for IDD-$m$, where $m$ represents the number of inversion pulses, is possible by giving one construction directly. By definition, CPC mitigates all control errors: it cancels all $k$-pulse echoes for any $1\le k\le m$, except for the desired echo. It is equivalent to constructing a matrix in which any subset of column vectors satisfies HPO. At present, we assume that the first $\pi/2$ pulse's phase is always $+1$ and focus on inversion pulses.

We construct a matrix encompassing all phase configurations of $m$ inversion pulses. Each pulse has two phase choices, $\pm1$. Accordingly, the matrix shape is $(2^{m},m)$. This matrix can be represented as:
\begin{equation}
	M= \{ g_{1}, g_{2}, g_{3}, \dots, g_m \}.
\end{equation}
We can easily check that any subset $K= \{ g_{k_{1}}, g_{k_{2}}, \dots, g_{k_{k}} \}$ of $M$ satisfies HPO, which requires half elements of the result vector $g_{i}=g_{k_{1}}\circ g_{k_{2}}\circ \dots \circ g_{k_{k}}$ equal to $+1$. Denote the number of $+1$ as $N(+1)$ and the number of $-1$ as $N(-1)$. The $k$ vectors give rise to $2^k$ phase configurations, which are repeated $2^{m-k}$ times in the subset $K$. Therefore, we only need to consider one of the repetitions. Imagine $k$ pulses whose phases are all initially $+1$. We can choose $0\le j\le k$ of them to become $-1$, which gives rise to $2^k$ possible constructions of row vectors of the subset K. If we chose an even number of the original row to become $-1$, the product of all elements within the resulting row would be $+1$. If we chose an odd number of the original row to become $-1$, the product would be $-1$. We thus have:
\begin{equation}
	\begin{aligned}
		N(+1) &= \sum_{even}C_{k}^{even},\\
		N(-1) &= \sum_{odd}C_{k}^{odd}.
	\end{aligned}
\end{equation}
$N(+1)$ equals to $N(-1)$ because:
\begin{equation}
	\begin{aligned}
		(-1+1)^{k}&=C_{k}^{0}(-1)^{k}+C_{k}^{1}(-1)^{k-1}+\dots+C_{k}^{k}(-1)^{0}\\
		&=(\sum_{even}C_{k}^{even}-\sum_{odd}C_{k}^{odd})(-1)^{k}=0
	\end{aligned}
\end{equation}
Therefore, any subset of $M$ satisfies HPO. Namely, $M$ represents CPC.
\\
\textbf{Complexity}:
While theoretically elegant, CPC becomes impractical for large $m$, requiring $2^m$ experiments. We prove that this is the minimal requirement.

Assume that $M'$ is a CPC matrix $(N,m)$ consisting of $\left \{g_{1},g_{2},\dots,g_{m}  \right \}$. For any $i\ne j$, $g_{i}\circ g_{j}\notin M'$, because if $g_{i}\circ g_{j}=g_{k}\in M'$, then $g_{i}\circ g_{j}\circ g_{k}=g_{k}\circ g_{k}=E$, which is inconsistent with the property of CPC. Hence, $g_{i}\circ g_{j}$ results in a new element. It is orthogonal to any element in $M'$ because $g_{k}\circ g_{i}=g_{j}$, $g_{k}\circ g_{j}=g_{i}$, and $g_{k}$ and any other element in $M'$ must fulfill HPO according to the property of CPC. We can construct $C_{m}^{2}$ such new elements. Similarly, they are orthogonal to each other. All new elements plus the original $M'$ form a matrix whose all column vectors are mutually orthogonal. For same reasons, we can also group $3,4,\dots,m$ elements together, and finally construct a much bigger matrix whose number of columns is:
\begin{equation}
	C_{m}^{1}+C_{m}^{2}+C_{m}^{3}+\dots+C_{m}^{m}=(1+1)^{m}-C_{m}^{0}=2^{m}-1
\end{equation}

Note that we can always add $E$ as the first column of the matrix, so finally we have $2^{m}$ elements. At this point, all the elements are still mutually orthogonal; equivalently, they are linearly independent. Therefore, the number of rows of the matrix (dimension of the present vector space) must be larger than or equal to the number of columns (dimension of the set of linearly independent vectors). Finally, we have:
\begin{equation}
	N\ge 2^{m}.
\end{equation}
The complexity of CPC is at least $2^{m}$. The above construction of CPC is optimal.
\\
\textbf{Experimental Consideration:} While theoretically setting the first pulse's phase to $E$ suffices, practical implementations often phase cycling the first pulse as $\begin{pmatrix}E \\ -E\end{pmatrix}$ to eliminate baseline drift. This is actually a combination of CPC and two-step phase cycling (TPC). Fortunately, we can now set the first inversion pulse's phase to be $E$ because it does not cause any unwanted signal if we neglect anti-echoes. So, the complexity is still $2^m$.
\\
\textbf{Two-step phase cycling:}
TPC involves two quantum circuits that differ only in phase of the first $\pi/2$ pulse; one uses a $\pi_x/2$ pulse and the other uses $\pi_{-x}/2$ pulse. Subtracting the results of these two circuits effectively cancels baseline drift. For experiments on single trapped ion, superconducting qubits, and NV centers in diamond, an extra $\pi/2$ pulse is applied before measuring population distribution. In this case, phase cycling the final $\pi/2$ pulse also suffices.

\subsubsection*{Hadamard phase cycling (HPC)}
Since CPC has exponential complexity $\mathcal{O}(2^{m})$, it is meaningful to design a good approximate phase cycling matrix that eliminates as many unwanted signals as possible while maintaining as low complexity as possible. Here, we show that regarding IDD-$m$ sequences, HPC has only linear complexity $\mathcal{O}(m)$ while canceling out more than $98\%$ unwanted signals for $m>16$.
\\
\textbf{Sylvester's construction of Hadamard matrix}:
Hadamard matrix is a square matrix of elements $+1$ and $-1$, where any two columns of the matrix are orthogonal: $H^{T}H=nI$.
Mathematicians suspect that any Hadamard matrix of order $4n$ always exists, where $n$ is an integer. This Hadamard conjecture remains unproved, but there exist many methods to construct Hadamard matrix. Here, we use Sylvester's method to construct Hadamard matrix of order $2^{n}$:

\begin{equation}
	H_{2^{n}}=
	\begin{pmatrix}
		H_{2^{n-1}} & H_{2^{n-1}} \\
		H_{2^{n-1}} & -H_{2^{n-1}}
	\end{pmatrix}, H_1=\begin{pmatrix}1
	\end{pmatrix}.
\end{equation}
Note that this matrix's first column is always an all-one vector, and due to its original orthogonality, it automatically mitigates all stimulated echoes that overlap with the desired echoes of CPMG.

\subsubsection*{Phase cycling group}
Represent Sylvester's Hadamard matrix as $H=\left \{ E,g_{1},g_{2},\dots,g_{H} \right \} $ like previous discussions. This set is an abelian group under Hadamard product, resulting in great HPO that means although not all subsets of the phase cycling matrix satisfy HPO, a significantly large ratio of subsets do satisfy.

Note that an abelian group requires the elements to satisfy these five properties:

\textbf{Closure:} $\forall g_{i},g_{j}\in H;g_{i}\circ g_{j}\in H.$

\textbf{Identity Existence:} $\forall g,\exists E;g\circ E=E\circ g=g.$

\textbf{Inverse Existence:} $\forall g,\exists g^{-1};g\circ g^{-1}=g^{-1}\circ g=E.$

\textbf{Associative law:} $\forall g_{i},g_{j},g_{k}\in H,(g_{i}\circ g_{j})\circ g_{k}=g_{i}\circ (g_{j}\circ g_{k}).$

\textbf{Commutative law:} $\forall g_{i},g_{j}\in H,g_{i}\circ g_{j}=g_{j}\circ g_{i}.$

Hadamard product is essentially a multiplication, so it automatically satisfies the associative law and the commutative law. The identity in $H$ is $E$, which consists of only $+1$ elements. The inverse of an element is just itself. Here, we prove the closure of $H$. First, notice that $H_{2}$ obviously satisfies the closure. Based on the Sylvester's construction of $H_{4}$ from $H_{2}$, the new elements in $H_{4}$ also satisfy the closure because they are formed by stacking elements in $H_{2}$. Similar logic is applicable to any higher-order matrix. In fact, one can treat $\begin{pmatrix}
	H & H \\
	H & -H
\end{pmatrix}$ as some kind of $H_{2}$, then the closure becomes obvious.
\begin{equation}
	H_{2}=
	\begin{pmatrix}
		1 & 1 \\
		1 & -1 \\
	\end{pmatrix}
	,
	H_{4}=
	\begin{pmatrix}
		1 & 1 & 1 & 1\\
		1 & -1 & 1 & -1\\
		1 & 1 & -1 & -1\\
		1 & -1 & -1 & 1
	\end{pmatrix}.
\end{equation}
\subsubsection*{Great Hadamard product orthogonality}
We now prove the great HPO of the phase cycling group. We first prove a conclusion to give some intuitions of the phase cycling group's great HPO:
\begin{equation}
	\begin{aligned}
		g_{1}\circ g_{2}\circ\dots \circ g_{i-1}\circ g_{i}&=E \\
		g_{1}\circ g_{2}\circ\dots \circ g_{i-1}&=E\circ g_{i} \\
		g_{1}\circ g_{2}\circ\dots \circ g_{i-1}\circ g_{j}&=E\circ g_{i}\circ g_{j} \\
		g_{i}\circ g_{j}=g_{k}&\neq E
	\end{aligned}
\end{equation}
Accordingly, if we had $i$ elements that do not satisfy HPO, we would be able to change any one of them to another element beyond the original set to construct a new set whose elements satisfy HPO.

We now discuss the proportion of the number of combinations of elements that do not satisfy HPO. Assume that we have $2^n$ elements in the phase cycling group $G_{H}$, and pick $q$ elements from $G_{H}-\left \{E  \right \} $. Define $p(q)$ as the probability that these elements do not satisfy HPO, and define $D(q)$ as the number of combinations satisfying $g_{1}\circ g_{2}\dots\circ g_{q}=E$, where $1\le q\le2^n-1$. We have:
\begin{equation}
	p(q)=\frac{D(q)}{C_{2^n-1}^{q}}.
\end{equation}
Due to the definition of $G_H$, $p(1)=p(2)=0$.

When $q=3$, we need the three elements to satisfy $g_3=g_{1}\circ g_{2}\neq E$. Given two elements, then the third element is uniquely determined. However, there are double counting issues while choosing the third element (we have to determine which one of the three elements is the third element), finally:
\begin{equation}
	D(3)=\frac{C_{2^n-1}^2}{C_3^1}.
\end{equation}

When $q=4$, we need the elements to satisfy $g_4=g_{1}\circ g_{2}\circ g_3\neq E$. Again, given three elements, the fourth element is uniquely determined. Moreover, we have to subtract $D(3)$ combinations satisfying $g_1\circ g_{2}\circ g_3=E$, because in this case, $g_4$ would have to be $E$ to satisfy our requirement, which is impossible. Taking the double counting issue into account, we have:

\begin{equation}
	D(4)=\frac{C_{2^n-1}^3 - D(3)}{C_4^1}.
\end{equation}

When $q=5$, we need the elements to satisfy $g_5=g_{1}\circ g_{2}\circ g_3\circ g_4\neq E$. Again, given four elements, the fifth element is uniquely determined. Furthermore, we have to subtract combinations satisfying $g_{1}\circ g_{2}\circ g_3\circ g_4=E$ for same reasons as $q=4$ case. In addition, when $g_{1}\circ g_{2}\circ g_3\circ g_4\neq E$, we need to subtract combinations satisfying $g_{1}\circ g_{2}\circ g_3=E$ because this leads to an impossible relation, $g_4\circ g_5=E$. Similarly, we have $D(4)$ combinations for the former one. For the latter one, we have $D(3)$ combinations for $g_1,g_2,g_3$, and $2^n-1-3$ choices for the fourth elements. Finally:
\begin{equation}
	D(5)=\frac{C_{2^n-1}^4 - D(4)-(2^n-1-3)D(3)}{C_5^1}.
\end{equation}
In fact, the situation of $q=5$ includes all necessary conditions. Hence, we can consider the general case, i.e. $q\ge5$ elements satisfying $g_{1}\circ g_{2}\circ\dots\circ g_q=E$. Given $q-1$ elements, we need to subtract $D(q-1)$ combinations for $g_{1}\circ g_{2}\circ\dots\circ g_{q-1}=E$, and then subtract $(2^n-1-q+2)D(q-2)$ combinations for $g_{1}\circ g_{2}\circ\dots\circ g_{q-2}=E$. We thus have:
\begin{equation}
	\begin{aligned}
		D(q)&=\frac{C_{2^n-1}^{q-1} - D(q-1)-(2^n-q+1)D(q-2)}{C_q^1},\\
		p(q)&=\frac{1-p(q-1)-(q-1)p(q-2)}{2^n-q}.
	\end{aligned}
\end{equation}
It is straightforward to check that only $q=1$ violates this relation, but it is a trivial case with $D(1)=p(1)=0$. Besides, $D(q)=D(2^n-1-q)$ because $g_1\circ g_2\circ\dots \circ g_{2^n-1}=E$.

We continue to derive the general formula for $D(q)$ and $p(q)$ based on the recursive formula. First, we prove a lemma:
\begin{equation}
	p(2k)=p(2k-1).
\end{equation}
When $k=1$, $p(2)=p(1)=0$. When $k=l$, assume that $p(2l)=p(2l-1)$. Hence, when $k=l+1$, using equation $(S28)$:
\begin{equation}
	\begin{aligned}
		(2^n-2l-1)\times p(2l+1)&=1-p(2l)-2l\times p(2l-1),\\
		(2^n-2l-2)\times p(2l+2)&=1-p(2l+1)-(2l+1)\times p(2l).
	\end{aligned}
\end{equation}
From these equations, we get:
\begin{equation}
	p(2(l+1))=p(2(l+1)-1).
\end{equation}
We thus prove the lemma.

For $q=2k+1$:
\begin{equation}
	\begin{aligned}
		D(2k+1)&=\frac{C_{2^n-1}^{q-1}-D(q-1)-(2^n-q+1)D(q-2)}{q}\\
		&=\frac{C_{2^n-1}^{q-1}}{q}-\frac{(2^n-q+1)D(q-2)}{q-1}.
	\end{aligned}
\end{equation}
Write $\frac{C_{2^n-1}^{q-1}}{q}$ as $\frac{C_{2^n-1}^q}{2^n}+\frac{(2^n-q+1)C_{2^n-1}^{q-2}}{(q-1)2^n}$, and change $q$ to $2k+1$:
\begin{equation}
	D(2k+1)-\frac{C_{2^n-1}^{2k+1}}{2^n}=-\frac{2^{n-1}-k}{k}\left [ D(2k-1)-\frac{C_{2^n-1}^{2k-1}}{2^{n}} \right ].
\end{equation}
Use this relation and $D(1)=0$, we finally get:
\begin{equation}
	D(2k+1)=\frac{C_{2^n-1}^{2k+1}}{2^n}+(-1)^{k+1}\times \frac{(2^n-1)C_{2^{n-1}-1}^{k}}{2^n}.
\end{equation}
Similarly:
\begin{equation}
	D(2k)=\frac{C_{2^n-1}^{2k}}{2^n}+(-1)^{k}\times \frac{(2^n-1)C_{2^{n-1}-1}^{k}}{2^n}.
\end{equation}
For $p(q)$, we have:
\begin{equation}
	\begin{aligned}
		p(2k)&=\frac{1}{2^n}+(-1)^{k}\times\frac{2^n-1}{2^n}\times\frac{C_{2^{n-1}-1}^k}{C_{2^{n}-1}^{2k}},\\
		p(2k+1)&=\frac{1}{2^n}+(-1)^{k+1}\times\frac{2^n-1}{2^n}\times\frac{C_{2^{n-1}-1}^k}{C_{2^{n}-1}^{2k+1}}.
	\end{aligned}
\end{equation}

Theoretically, for an IDD sequence with 32 inversion pulses, CPC requires $2^{32}=4,294,967,296$ steps, while the phase cycling based on Hadamard matrix only requires $32$ steps. The HPO ratio, which is defined as $1-\sum_{k}\frac{D(k)}{C_{2^n-1}^{k}}$, is $96.88\%$.
\subsubsection*{Final phase cycling scheme}
We further improve the HPO ratio by harnessing the parity of coherence transfer pathways and TPC. Note that the Sylvester's Hadamard matrix has extra freedoms to create new Hadamard matrices. For example, if we change the first column from $E$ to $-E$, the matrix is still a Hadamard matrix with great HPO. We can also add a minus sign to all the rest columns as well, i.e. changing $H'\to-H'$, to construct a new stacked Hadamard matrix:
\begin{equation}
	\begin{pmatrix}
		E & H'\\
		E & -H'\\
	\end{pmatrix}.
\end{equation}
The advantage of this new matrix is that any odd number of elements excluding the first column satisfy HPO. Formally, if $q$ is an odd number, $D(q)=0$. Imagine that there are odd number of elements in $H'$ forming an unwanted signal; namely, these elements do not satisfy HPO. The Hadamard product of these elements in $-H'$ has an opposite sign. Thus, adding these two Hadamard products together would satisfy HPO. Such stacking increases the HPO ratio from $96.88\%$ to $98.44\%$ for IDD-32.
Finally, we can add a column for the first $\pi/2$ pulse to construct HPC:
\begin{equation}
	\begin{pmatrix}
		+ & E & E & H'\\
		+ & E & E & -H'\\
		- & -E & E & H'\\
		- & -E & E & -H'
	\end{pmatrix}.
\end{equation}
Note that we do not need the first two columns to satisfy HPO if we neglect anti-echoes. For IDD-$m$, we need $4\times N_{HPC}$ phase cycling steps, where $N_{HPC}=2^{integer}$ is the smallest number satisfying $N_{HPC}\ge m$. For example, HPC needs $128$ phase cycling steps for IDD-32, but it needs $256$ phase cycling steps for IDD-34. In sharp contrast, CPC would need $4,294,967,296$ and $17,179,869,184$ steps for IDD-32 and IDD-34, respectively, both of which are clearly not practical.

Overall, whereas CPC has a complexity of $\mathcal{O}(2^m)$ for IDD-$m$, HPC significantly reduces the complexity to $\mathcal{O}(m)$ while preserving near-quantitative HPO ratio. HPC is both scalable and effective.

\subsubsection*{Effective State Fidelity}
\textbf{Definition}:
State fidelity serves as a merit of how close two quantum states are, which is typically defined as:
\begin{equation}
	F(\rho,\sigma)=(Tr \sqrt{\sqrt{\rho} \sigma\sqrt{\rho}})^2.
\end{equation}
Here, $\sigma$ represents the target state and $\rho$ the state for comparison. We define:
\begin{equation}
	\begin{aligned}
		S_{x,y,z} &= \sum_i V_{s}^is_{x,y,z}^i \\
		\left \langle S_{x,y,z} \right \rangle &=  \frac{S_{x,y,z}}{\sqrt{S_x^2+S_y^2+S_z^2}},
	\end{aligned}
\end{equation}
where $V_{s}^i$ is the $i_{th}$ element of the sign vector (see equation (S38)) and $s_{x,y,z}^i$ are expectation values of Pauli operators for the $i_{th}$ phase cycling step. We call $\left \langle S_{x,y,z} \right \rangle$ as phase-cycled expectation values and use them to reconstruct the effective quantum state:
\begin{equation}
	\rho^{eff}=\frac{1}{2}(\hat{S}_{0}
	+ \left \langle S_x \right \rangle \hat{S}_{x}
	+ \left \langle S_y \right \rangle \hat{S}_{y}
	+ \left \langle S_z \right \rangle \hat{S}_{z}).
\end{equation}
Finally, the effective state fidelity reads
\begin{equation}
	F(\rho^{eff},\sigma)=(Tr \sqrt{\sqrt{\rho^{eff}} \sigma\sqrt{\rho^{eff}}})^2.
\end{equation}
The effective state fidelity serves as a useful benchmark for phase cycling. Essentially, phase cycling sums different Bloch vectors, which represent qubit states, to get a more accurate Bloch vector. Thus, the better the phase cycling method, the higher the effective state fidelity (if the DD sequence does not have inherent robustness).

\subsubsection*{Illustrative coherent pulse-error model}
\label{sec:coherent-pulse-error}
To examine the plausibility of coherent error accumulation in the trapped-ion fidelity measurements, we simulated inversions $R_x[q(\pi+\epsilon)]$, where $R_x(\theta)=\exp(-i\theta\sigma_x/2)$, $\sigma_x$ is the Pauli matrix, $q=\pm1$ specifies opposite pulse phases, and $\epsilon$ is a fixed rotation-angle error. The transverse input was perpendicular to the rotation axis; preparation and readout were ideal, and free evolution was the identity in the rotating frame. An illustrative $1.2\%$ pulse-area error, $\epsilon=0.012\pi=2.16^\circ$, gives an average single-pulse gate fidelity $F_{\rm avg}=1-\frac{2}{3}\sin^2(\epsilon/2)=99.9763\%$. After 64 same-axis pulses, however, $F_{\rm eff}^{\rm TPC}=\cos^2(64\epsilon/2)=0.127029$, whereas HPC gives $F_{\rm eff}^{\rm HPC}=0.999645$ with the same pulse error. This example reproduces the scale of the observed fidelity collapse without fitting the experimental data or identifying a unique microscopic error source. The ideal-free-evolution model does not distinguish CP from UDD.

\subsubsection*{Control-error comparison of XY8 and XY16}
\label{sec:robust-control-comparison}
In unit-level phase cycling, each complete XY8 or XY16 block is treated as one pulse unit, and all pulses within the block receive a common Hadamard-selected $0$ or $\pi$ phase shift. This preserves the internal relative phases. Pulse-level phase cycling instead assigns Hadamard phase shifts to individual pulses, which can disrupt the sequence's error compensation.

To compare these schemes, we simulated a single qubit driven by rectangular nominal $\pi$ pulses of duration $\pi/\Omega_0$, where $\Omega_0$ is the nominal Rabi angular frequency. Preparation and readout were ideal; free evolution was the identity, with no environmental decoherence or longitudinal relaxation. Each sequence contained 128 XY8 or XY16 blocks (1024 or 2048 pulses, respectively). XY8 used $XYXYYXYX$; XY16 appended a copy with all phases shifted by $\pi$. Pulse-amplitude deviations and detunings were constant within each realization and scanned over $\pm10\%$ of $\Omega_0$. We compared TPC, unit-level HPC, and pulse-level HPC over these error ranges using the average gate fidelity relative to ideal DD evolution, averaged uniformly over pure input states.

Figure~\mbox{\ref{fig:xy8-xy16-control}} shows that unit-level HPC expands the XY8 region with $F_{\rm avg}>0.95$, whereas TPC and unit-level HPC give nearly identical XY16 results (maximum absolute difference below $4.9\times10^{-5}$). Pulse-level HPC leaves larger low-fidelity regions near parts of the scan boundaries. In quantum sensing, finite pulse durations and detuning can nevertheless produce spurious responses in XY8/XY16 protocols\mbox{\cite{si_spurious}}. The related Hadamard sensing channel (HSC) suppresses these artifacts in XY8 simulations; its weak-coupling theory also supports applying the approach to repeated XY16 blocks\mbox{\cite{si_HadamardSensingChannel}}.

\begin{figure}
\centering
\includegraphics[width=0.6\textwidth]{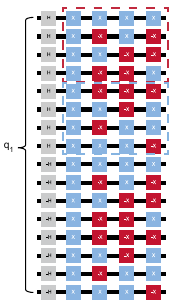}

\caption{\textbf{Hadamard phase cycling for IDD-4.}
q$_1$ and H represent qubit and Hadamard gate ($\pi/2$ gate), respectively. $\pm X$ represent inversion pulses with opposite phases. Phase configurations within red block represent the original Hadamard matrix, and phase configurations within blue block represent a minus Hadamard matrix except for the first column.  }
\label{fig:s1}
\end{figure}

\clearpage
\begin{figure}
\centering
\includegraphics[width=\textwidth]{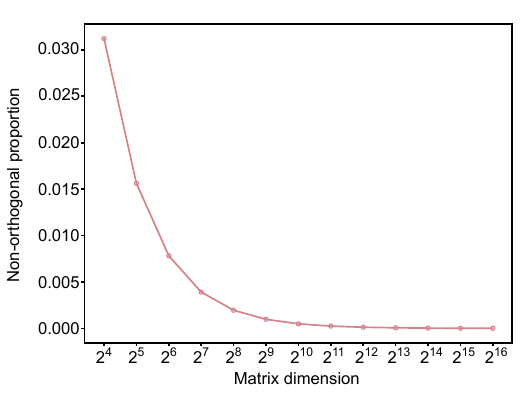}

\caption{\textbf{Non-orthogonal proportion for HPC.}
Non-orthogonal ratio ($1-$HPO ratio) of HPC with different dimensions of Hadamard matrices.}
\label{fig:s2}
\end{figure}

\clearpage
\begin{figure}
\centering
\includegraphics[width=\textwidth]{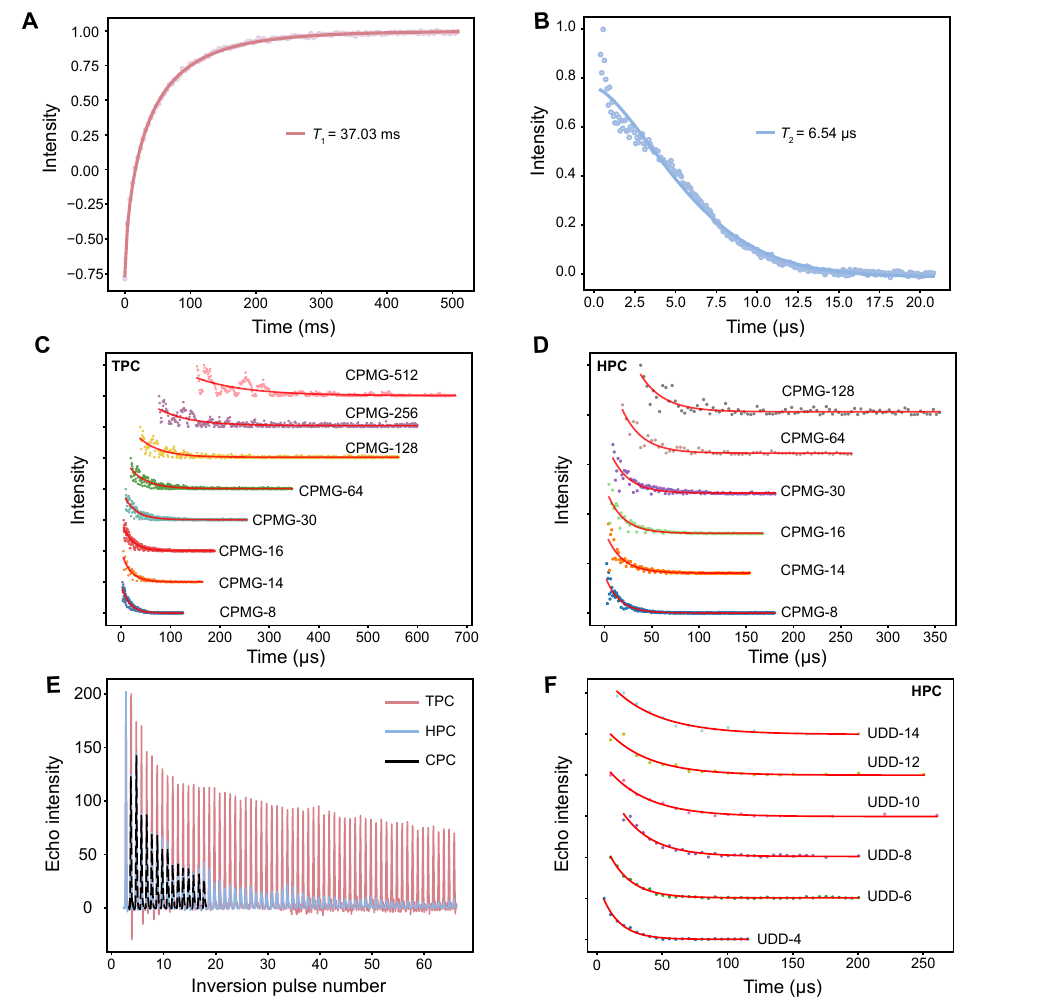}

\caption{\textbf{Pulse EPR spectroscopic characterization results of (PPh$_\textbf{4}$)$_\textbf{2}$[Cu(mnt)$_\textbf{2}$] at 8 K.}
\textbf{A}, inversion recovery curve. \textbf{B}, Hahn echo decay curve. \textbf{C,D}, echo decay curves for CPMG experiments with TPC and HPC, respectively. \textbf{E}, comparison between echoes collected under CPMG with different phase cycling methods. \textbf{F}, echo decay curves for UDD experiments with HPC.}
\label{fig:s3}
\end{figure}

\clearpage
\begin{figure}
\centering
\includegraphics[width=\textwidth]{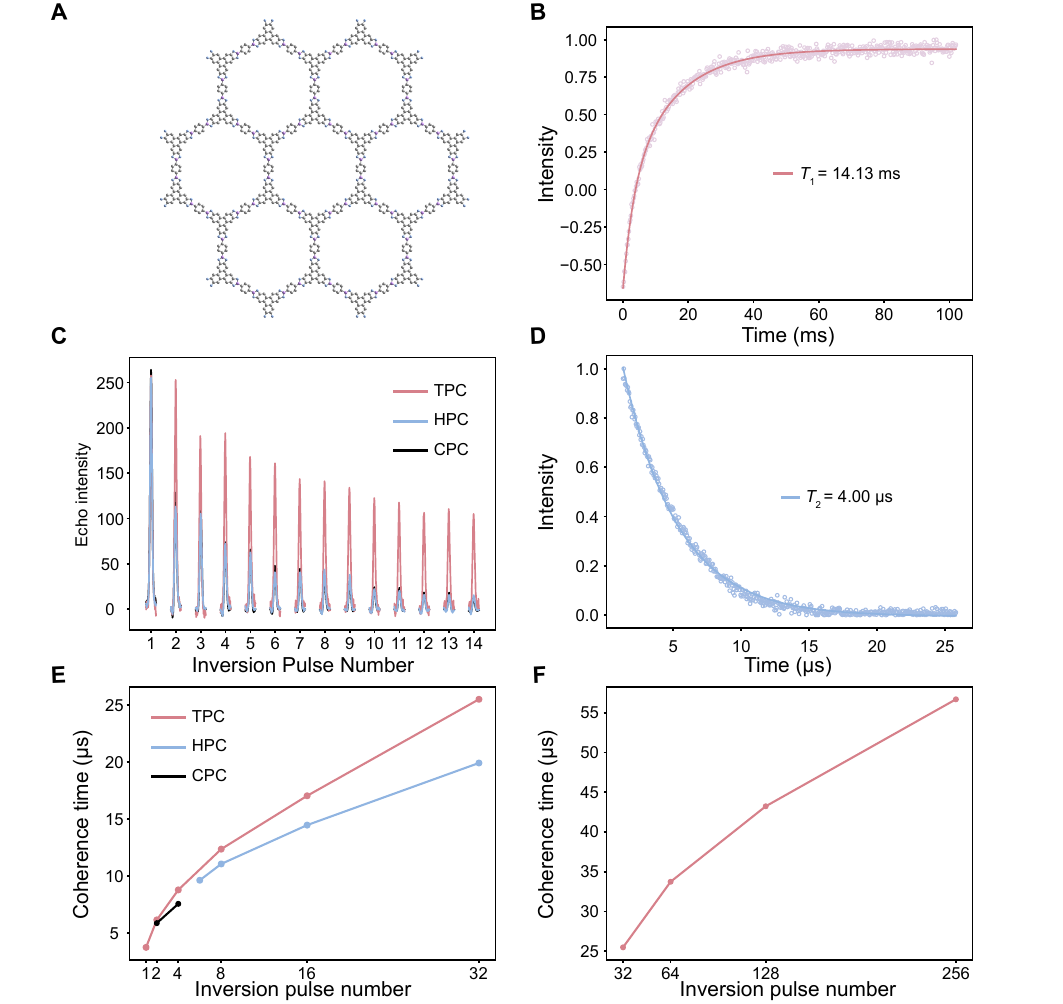}

\caption{\textbf{Pulse EPR spectroscopic characterization results of COF-5 at 80 K.}
\textbf{A}, structure of COF-5. \textbf{B}, inversion recovery curve. \textbf{C}, comparison between echoes collected under CPMG with different phase cycling methods. \textbf{D}, Hahn echo decay curve. \textbf{E}, comparison between $T_2$ measured by CPMG with different phase cycling methods. \textbf{F}, $T_2$ measured by CPMG with TPC.}
\label{fig:s4}
\end{figure}

\clearpage
\begin{figure}
\centering
\includegraphics[width=\textwidth]{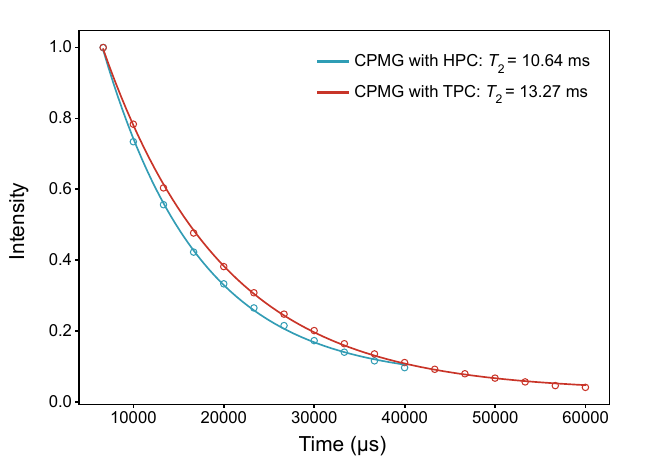}

\caption{\textbf{CPMG Results with different phase cycling methods of SSNMR.}}
\label{fig:s5}
\end{figure}

\clearpage
\begin{figure}
\centering
\includegraphics[width=\textwidth]{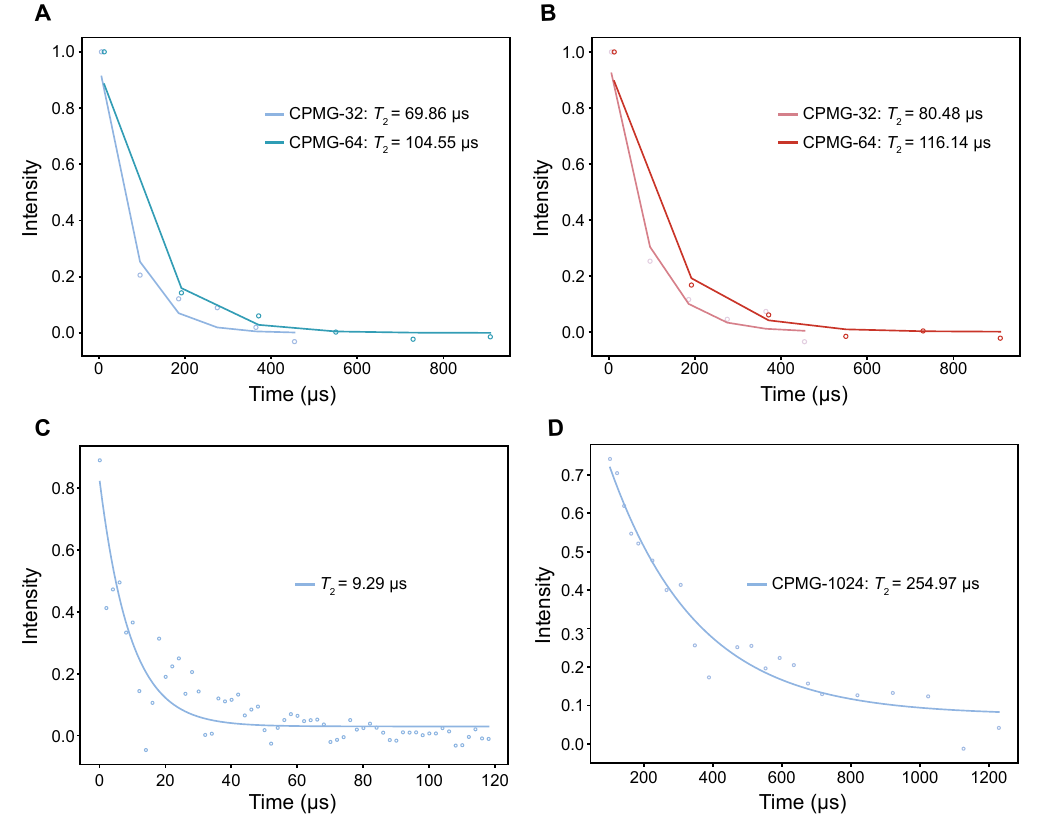}

\caption{\textbf{Optically detected magnetic resonance spectroscopic characterization results of diamond NV Centers.}
\textbf{A,B}, signal decay curves of CPMG-32 and CPMG-64 with HPC and TPC, respectively. \textbf{C}, Hahn echo decay curve. \textbf{D}, echo decay curve collected by CPMG-1024 with TPC.}
\label{fig:s6}
\end{figure}

\clearpage
\begin{figure}
\centering
\includegraphics[width=\textwidth]{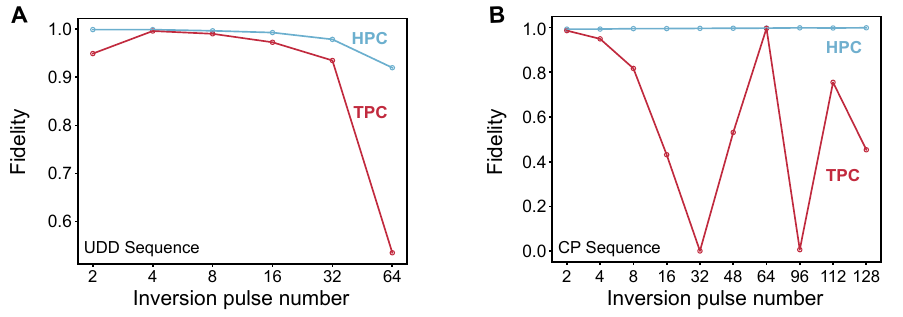}

\caption{\textbf{Effective state-fidelity tests on single superconducting transmon qubit.}
\textbf{A}, results for UDD experiments with TPC and HPC. \textbf{B}, results for CP experiments with TPC and HPC. A $\pi/28$ systematic error was introduced to each inversion pulse.}
\label{fig:s7}
\end{figure}

\clearpage
\begin{figure}
\centering
\includegraphics[width=\textwidth]{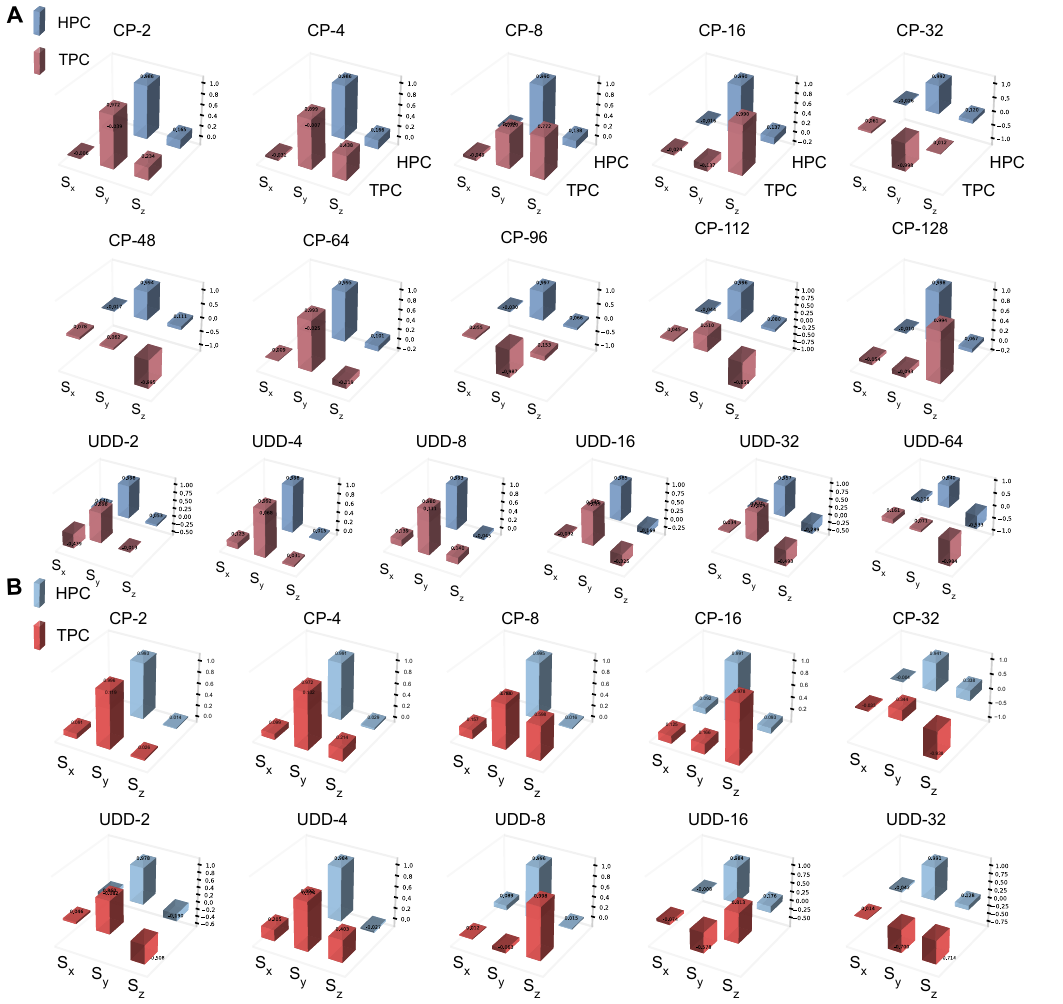}

\caption{\textbf{ Quantum state tomography results.}
\textbf{A}, Single superconducting transmon qubit results and \textbf{B}, single trapped ion qubit results.}
\label{fig:s8}
\end{figure}

\clearpage
\begin{figure}[p]
\centering
\makebox[\textwidth][c]{\includegraphics[width=6.5in]{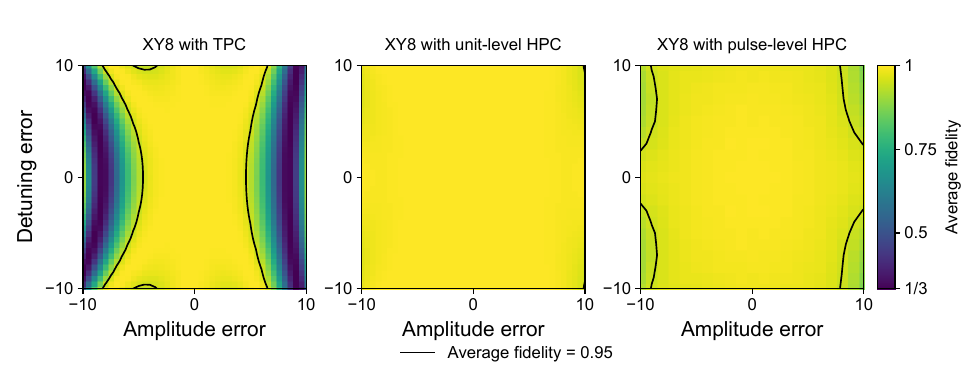}}
\par\vspace{0.15in}
\makebox[\textwidth][c]{\includegraphics[width=6.5in]{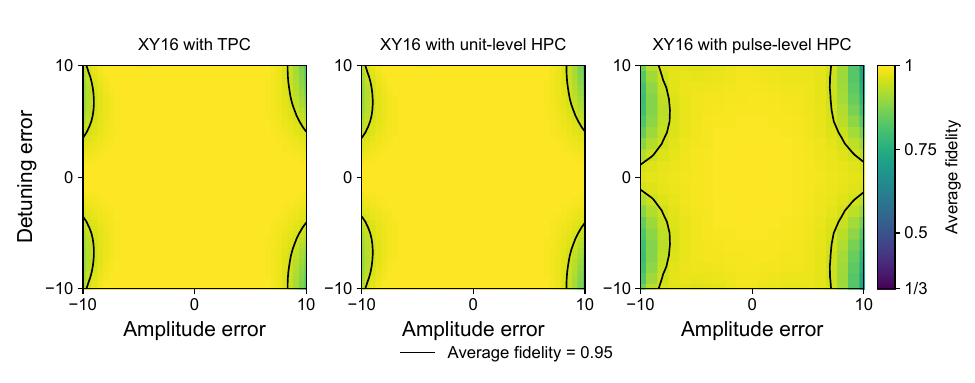}}
\caption{\textbf{Control-error comparison of XY8 and XY16.} Average gate fidelity for 128 complete XY8 blocks (top; 1024 inversion pulses) and XY16 blocks (bottom; 2048 inversion pulses). Columns compare TPC, unit-level HPC, and pulse-level HPC. Both error axes are percentages of the nominal Rabi frequency. Black contours mark $F_{\rm avg}=0.95$. TPC and unit-level HPC use $41\times41$ grids; pulse-level HPC uses a $21\times21$ grid. Unit-level HPC improves XY8 over much of the scanned range; for XY16, its maximum absolute difference from TPC is below $4.9\times10^{-5}$.}
\label{fig:xy8-xy16-control}
\end{figure}

\clearpage
\begin{table}
\centering

\caption{\textbf{Coherence transfer pathways in CPMG-2.}
Echo position is with respect to the last pulse.}
\label{tab:S1}

\begin{tabular}{ccccc}
\\
\hline
\textbf{$p_0$} & \textbf{$p_1$} & \textbf{$p_2$} & \textbf{$p_3$} & Echo Position\\
\hline
0 & $-1$ & $+1$ & $-1$ & $\tau$ \\
0 & $+1$ & $+1$ & $-1$ & $3\tau$\\
0 & 0 & $+1$ & $-1$ & $2\tau$ \\
0 & $+1$ & 0 & $-1$ & $\tau$ \\
\hline
\end{tabular}
\end{table}

\begin{table}
\centering

\caption{\textbf{Changes of coherence orders in CPMG-2}}
\label{tab:S2}

\begin{tabular}{ccccc}
\\
\hline
\textbf{$\Delta p_0$} & \textbf{$\Delta p_1$} & \textbf{$\Delta p_2$} & \textbf{$\Delta p_3$} & Echo Position \\
\hline
$None$ & $-1$ & $+2$ & $-2$ & $\tau$ \\
$None$ & $+1$ & $-1$ & $-1$ & $\tau$ \\
\hline
\end{tabular}
\end{table}

\begin{table}
\centering

\caption{\textbf{Complete phase cycling for CPMG-2.}
The symbol in the ``Sign'' column indicates addition or subtraction of the final signal. ``Original'' and ``Flipped'' represent phases of echoes.}
\label{tab:S3}

\begin{tabular}{cccccc}
\\
\hline
Sign & Pulse 1 & Pulse 2 & Pulse 3 & Desired echo & Stimulated Echo \\
\hline
$+$ & $+X$ & $+Y$ & $+Y$ & Original & Original \\
$+$ & $+X$ & $+Y$ & $-Y$ & Original & Flipped \\
$+$ & $+X$ & $-Y$ & $+Y$ & Original & Flipped \\
$+$ & $+X$ & $-Y$ & $-Y$ & Original & Original \\
\hline
\end{tabular}
\end{table}

\clearpage
\begin{table}
\centering

\caption{\textbf{Example of CPC for CPMG-4.}
Note: Consider a subset $K= \{g_3, g_4 \}$. Regarding the last two pulses, the first four rows already contain all possible phase configurations, and the rest of rows are simply their repetitions.}
\label{tab:S4}

\begin{tabular}{cccc}
\\
\hline
Pulse 1 & Pulse 2 & Pulse 3 & Pulse 4 \\
\hline
$+$ & $+$ & $+$ & $+$ \\
$+$ & $+$ & $+$ & $-$  \\
$+$ & $+$ & $-$ & $+$  \\
$+$ & $+$ & $-$ & $-$  \\

$+$ & $-$ & $+$ & $+$ \\
$+$ & $-$ & $+$ & $-$  \\
$+$ & $-$ & $-$ & $+$  \\
$+$ & $-$ & $-$ & $-$  \\

$-$ & $+$ & $+$ & $+$ \\
$-$ & $+$ & $+$ & $-$  \\
$-$ & $+$ & $-$ & $+$  \\
$-$ & $+$ & $-$ & $-$  \\

$-$ & $-$ & $+$ & $+$ \\
$-$ & $-$ & $+$ & $-$  \\
$-$ & $-$ & $-$ & $+$  \\
$-$ & $-$ & $-$ & $-$  \\
\hline
\end{tabular}
\end{table}

\clearpage
\begin{table}
\centering

\caption{\textbf{$\textbf{T}_\textbf{2}$ of (PPh$_\textbf{4}$)$_\textbf{2}$[Cu(mnt)$_\textbf{2}$] under IDD-$\textbf{m}$.}}
\label{tab:S5}

\begin{tabular}{cccc}
\\
\hline
$m$ & Phase cycling & $T_2~(\mathrm{\mu s})$ & IDD\\
\hline
8 &  HPC &       14.85 &        CPMG  \\
8 &  TPC &      16.80 &       CPMG \\
14 &  HPC &      17.17  &      CPMG  \\
14 &  TPC &    18.50 &     CPMG  \\
16 &  HPC &   17.39 &    CPMG  \\
16 &  TPC &   20.58 &    CPMG  \\
30 &  HPC &  18.80 &   CPMG  \\
30 &  TPC &  24.63 &   CPMG  \\
64 &  HPC & 21.13 &  CPMG  \\
64 & TPC & 33.13 &  CPMG  \\
128 & HPC & 24.42 &  CPMG  \\
128 & TPC & 43.53 & CPMG  \\
256 & TPC & 64.05 & CPMG  \\
512 & TPC & 98.91 & CPMG \\
4 & HPC & 13.70 & UDD  \\
6 & HPC & 16.88 & UDD  \\
8 & HPC & 23.22 & UDD  \\
10 & HPC & 30.56 & UDD  \\
12 & HPC & 31.56 & UDD  \\
14 & HPC & 35.72 & UDD  \\
\hline
\end{tabular}
\end{table}

\clearpage
\begin{table}
\centering

\caption{\textbf{$\textbf{T}_\textbf{2}$ of COF-5 under IDD-$\textbf{m}$}}
\label{tab:S6}

\begin{tabular}{cccc}
	\\
	\hline
	$m$ & Phase cycling & $T_2~(\mathrm{\mu s})$ & IDD\\
	\hline
	2 &  CPC &      5.88  &      CPMG  \\
	4 &  CPC &    7.57 &     CPMG  \\
	6 &  HPC &   9.63 &    CPMG  \\
	8 &  HPC &   11.06 &    CPMG  \\
	16 &  HPC &  14.47 &   CPMG  \\
	32 &  HPC &  19.91 &   CPMG  \\
	2 &  TPC & 6.16 &  CPMG  \\
	4 & TPC & 8.79 &  CPMG  \\
	8 & TPC & 12.36 &  CPMG  \\
	16 & TPC & 17.03 & CPMG  \\
	32 & TPC & 25.48 & CPMG  \\
	64 & TPC & 33.74 & CPMG \\
	128 & TPC & 43.23 & CPMG \\
	256 & TPC & 56.70 & CPMG \\
	\hline
\end{tabular}
\end{table}
\clearpage
\begin{table}[p]
\centering
\caption{\textbf{Normalized residual RMS of mono-exponential fits for (PPh$_4$)$_2$[Cu(mnt)$_2$] at 8 K.} The CPMG and UDD decays correspond to Fig.~\mbox{\ref{fig:s3}C,D,F} and the $T_2$ values in Table~\mbox{\ref{tab:S5}}. $N$ is the number of recorded points; residuals are defined under ``CPMG measurements''. Fits used $f(t)=C+A\exp(-t/T_2)$ with adjustable amplitude $A\ge0$, offset $C$, and $T_2$, determined by unweighted nonlinear least squares using all recorded points.}
\label{tab:mono-fit-residuals}
\begingroup
\fontsize{10}{12}\selectfont
\setlength{\tabcolsep}{6pt}
\renewcommand{\arraystretch}{1.2}
\begin{tabular*}{\textwidth}{@{\extracolsep{\fill}}llrrr@{}}
\toprule
Sequence & Phase cycling & $m$ & $N$ & Normalized RMS \\
\midrule
CPMG & HPC & 8 & 256 & 0.048022 \\
CPMG & HPC & 14 & 128 & 0.077696 \\
CPMG & HPC & 16 & 128 & 0.067426 \\
CPMG & HPC & 30 & 128 & 0.074976 \\
CPMG & HPC & 64 & 64 & 0.065308 \\
CPMG & HPC & 128 & 96 & 0.090040 \\
\midrule
CPMG & TPC & 8 & 512 & 0.051219 \\
CPMG & TPC & 14 & 128 & 0.061445 \\
CPMG & TPC & 16 & 512 & 0.058115 \\
CPMG & TPC & 30 & 512 & 0.057669 \\
CPMG & TPC & 64 & 512 & 0.061741 \\
CPMG & TPC & 128 & 512 & 0.069371 \\
CPMG & TPC & 256 & 512 & 0.098354 \\
CPMG & TPC & 512 & 512 & 0.123878 \\
\midrule
UDD & HPC & 4 & 23 & 0.023394 \\
UDD & HPC & 6 & 39 & 0.021060 \\
UDD & HPC & 8 & 32 & 0.049164 \\
UDD & HPC & 10 & 18 & 0.040975 \\
UDD & HPC & 12 & 18 & 0.079622 \\
UDD & HPC & 14 & 13 & 0.061956 \\
\bottomrule
\end{tabular*}
\endgroup
\end{table}

\clearpage
\endgroup
\end{document}